\documentstyle[aps,eqsecnum,twocolumn,amsfonts,amssymb,prd,epsfig,amsmath]{revtex}

\begin{document}
\title{Stochastic Theory of Relativistic Particles Moving in a Quantum Field: II.
Scalar Abraham-Lorentz-Dirac-Langevin Equation, Radiation Reaction and
Vacuum Fluctuations}
\author{Philip R. Johnson\thanks{%
Electronic address: {\tt pj19@umail.umd.edu}} and B. L. Hu\thanks{%
Electronic address: {\tt hub@physics.umd.edu}}}
\address{Department of Physics, University of Maryland,\\
College Park, Maryland 20742-4111}
\date{April 26, 2000}
\maketitle

\begin{abstract}
We apply the open systems concept and the influence functional formalism
introduced in Paper I to establish a stochastic theory of relativistic
moving spinless particles in a quantum scalar field. The stochastic regime
resting between the quantum and semi-classical captures the statistical
mechanical attributes of the full theory. Applying the particle-centric
world-line quantization formulation to the quantum field theory of scalar
QED we derive a time-dependent (scalar) Abraham-Lorentz-Dirac (ALD) equation
and show that it is the correct semiclassical limit for nonlinear
particle-field systems without the need of making the dipole or
non-relativistic approximations. Progressing to the stochastic regime, we
derive multiparticle time-dependent ALD-Langevin equations for nonlinearly
coupled particle-field systems. With these equations we show how to address
time-dependent dissipation/noise/renormalization in the semiclassical and
stochastic limits of QED. We clarify the relation of radiation reaction,
quantum dissipation and vacuum fluctuations and the role that initial
conditions may play in producing non-Lorentz invariant noise. We emphasize
the fundamental role of decoherence in reaching the semiclassical limit,
which also suggests the correct way to think about the issues of runaway
solutions and preacceleration from the presence of third derivative terms in
the ALD equation. We show that the semiclassical self-consistent solutions
obtained in this way are ``paradox'' and pathology free both technically and
conceptually. This self-consistent treatment serves as a new platform for
investigations into problems related to relativistic moving charges.
\end{abstract}

\section{Introduction}

This is the second in a series of three papers \cite{JH1,JH3} exploring the
regime of stochastic behavior manifested by relativistic particles moving
through quantum fields. This series highlights the intimate connections
between backreaction, dissipation, noise, and correlation, focusing on the
necessity (and the consequences) of a self-consistent treatment of
nonlinearly interacting quantum dynamical particle-field systems.

In Paper I \cite{JH1}, we have set up the basic framework built on the
influence functional and worldline quantization formalism. In this paper we
apply the results obtained there to spinless relativistic moving particles
in a quantum scalar field. The interaction is chosen to be the scalar analog
of QED coupling so that we can avoid the complications of photon
polarizations and gauge invariance in an electromagnetic field. The results
here should describe correctly particle motion when spin and photon
polarization are unimportant, and when the particle is sufficiently
decohered such that its quantum fluctuations effectively produce stochastic
dynamics.

The main results of this investigation are

\begin{enumerate}
\item  First principle derivation of a time-dependent (modified)
Abraham-Lorentz-Dirac (ALD) equation as the consistent semiclassical limit
for nonlinear particle-scalar-field systems without making the dipole or
non-relativistic approximations.

\item  Consistent resolution of the paradoxes of the ALD equations,
including the problems of runaway and acausal (e.g. pre-accelerating)
solutions, and other ``pathologies''. We show how the non-Markovian nature
of the quantum particle open-system enforces causality in the equations of
motion. We also discuss the crucial conceptual role that decoherence plays
in understanding these problems.

\item  Derivation of multiparticle Abraham-Lorentz-Dirac-Langevin (ALDL)
equations describing the quantum stochastic dynamics of relativistic
particles. The familiar classical Abraham-Lorentz-Dirac (ALD) equation is
reached as its {\it noise-averaged } form. The stochastic regime,
characterized by balanced noise and dissipation, plays a crucial role in
bridging the gap between quantum and (emergent) classical behaviors. The
N-particle-irreducible (NPI) and master effective action \cite
{CH93BrillVolume,CalzettaHu95cddn,CA96,CH00stobol} provide a route for
generalizing our treatment to the {\it self-consistent} inclusion of higher
order quantum corrections.
\end{enumerate}

We divide our introduction into three parts: First, a discussion of the
pathologies and paradoxes of the ALD equation from the conventional approach
and our suggested cures based on proper treatment of causal and
non-Markovian behavior and self-consistent backreaction. Second,
misconceptions in the relation between radiation reaction, quantum
dissipation and vacuum fluctuations and the misplaced nature and role of
fluctuation-dissipation relations. Finally we give a brief summary of
previous work and describe their shortcomings which justify a new approach
as detailed here.

\subsection{Pathologies, Paradoxes, Remedies and Resolutions}

The classical theory of moving charges interacting with a classical
electromagnetic field has controversial difficulties associated with
backreaction \cite{Pathologies,MilonniQVacuum94}. The generally accepted
classical equation of motion in a covariant form for charged, spinless point
particles, including radiation-reaction, is the Abraham-Lorentz-Dirac (ALD)
equation \cite{ALD}: 
\begin{equation}
\ddot{z}^{\mu}+\left( 2e^{2}/3m\right) (\dot{z}^{\mu}\ddot{z}^{2}+\stackrel{%
...}{z}^{\mu})=(e/m)\dot{z}_{\nu}F_{ext}^{\mu\nu}(z).
\label{Classical Lorentz-Dirac equation}
\end{equation}
The timescale $\tau_{0}=\left( 2e^{2}/3mc^{3}\right) $ determines the
relative importance of the radiation-reaction term. For electrons, $\tau
_{0}\sim10^{-24}$ secs, which is roughly the time it takes light to cross
the electron classical radius $r_{0}\sim10^{-15}$m. The ALD equation has
been derived in a variety of ways, often involving some regularization
procedure that renormalizes the particle's mass \cite
{Pearle82(EM:Paths.to.research)}. Feynman and Wheeler have derived this
result from their ``Absorber'' theory which symmetrically treats both
advanced and retarded radiation on the same footing \cite
{FeynmanWheeler45(Absorber.Theory)}.

The ALD equation has strange features, whose status are still debated.
Because it is a third order differential equation, it requires the
specification of extra initial data (e.g. the initial acceleration) in
addition to the usual position and velocity required by first order
Hamiltonian systems. This leads to the existence of runaway solutions.
Physical (e.g. non-runaway) solutions may be enforced by transforming (\ref
{Classical Lorentz-Dirac equation}) to a second order integral equation with
boundary condition such that the final energy of the particle is finite and
consistent with the total work done on it by external forces. But the
removal of runaway solutions comes at a price, because solutions to the
integral equation exhibit the acausal phenomena of pre-acceleration on
timescales $\tau_{c}.$ This is the source of lingering questions on whether
the classical theory of point particles and fields is causal\footnote{%
Another mystery of the ALD equation is the problem of the constant force
solution, where the charge uniformly accelerates without radiation reaction,
despite it being well-established that there is radiation. This is a case
where local notions of energy must be carefully considered in theories with
local interactions (between point particles) mediated by fields \cite
{Rohrlich.Fulton60:Class.Rad.uniformlly.acc.charge,Rohrlich69(Rad.Reaction.Charged.Particle)}%
. The classical resolution of this problem involves recognizing that one
must consider all parts of the total particle-field system with regard to
energy conservation: particle energy, radiation energy, and non-radiant
field energy such as resides in the so-called acceleration (or Shott) fields 
\cite{Rohrlich69(Rad.Reaction.Charged.Particle)}.}.

There have been notable efforts to understand charged particle
radiation-reaction in the classical and quantum theory (for a review see 
\cite{MilonniQVacuum94}), of which extended charge theories \cite
{ExtendedCharges} and quantum Langevin equations are important examples \cite
{QLEs,CaldeiraLeggett,FOL} (See Subsection C below).

\subsection{Radiation Reaction and Vacuum Fluctuations}

The stochastic regime is characterized by competition between quantum and
statistical processes: particularly, quantum correlations versus
decoherence, and fluctuations versus dissipation. The close association of
vacuum fluctuations with radiation-reaction is well-known, but the precise
relationship between them requires clarification. It is commonly asserted
that vacuum fluctuations (VF) are balanced by radiation reaction (RR)
through a fluctuation-dissipation relation (FDR). This can be misleading.
Radiation reaction is a classical process (i.e., $\hslash=0$) while vacuum
fluctuations are quantum in nature. A counter-example to the claim that
there is a direct link between RR and VF is the uniformly accelerated
charged particle (UAP) for which the classical radiation reaction force
vanishes, but vacuum fluctuations do not. However, at the stochastic level
variations in the radiation reaction force away from the classical averaged
value exist -- it is this quantum dissipation effect which is related to
vacuum fluctuations by a FDR.

From discussions in Paper I we see how decoherence enters in the quantum to
classical transition. Under reasonable physical conditions \cite{GHDK} the
influence action or decoherence functional is dominated by the classical
solution (via the stationary phase approximation). Also from the
`non-Markovian' fluctuation-dissipation relations found in Paper I, we see
that the relationship between noise and dissipation is extremely
complicated. Here too, decoherence plays a role in the emergence of the
usual type fluctuation-dissipation kernel, just as it plays a role in the
emergence of classicality. When there is sufficient decoherence, the FDR
kernel (see Eq. (5.23) in Paper I) is dominated by the self-consistent
semiclassical (decoherent) trajectories. An approximate, linear FDR relation
may be obtained (see Eq. (5.24) in Paper I) except that the kernel is
self-consistently determined in terms of the average (mean) system history,
and the FDR describes the balance of fluctuations and dissipation about
those mean trajectories.

These observations have important ramifications on the appropriateness of
assigning a FDR for radiation reaction and vacuum fluctuations. Ordinary
(e.g. classical) radiation reaction is consistent with the mean particle
trajectories determined by including the backreaction force from the field.
But at the classical level, there are no fluctuations, and hence no
fluctuation-dissipation relation. Even regarding radiation reaction as a
`damping' mechanism is misleading-- it is not `dissipation' in the true
statistical mechanical sense. Unlike a particle moving in some viscous
medium whose velocity is damped, radiation reaction generally vanishes for
inertial (e.g. constant velocity) particles moving in vacuum fields (e.g.
QED). Furthermore, radiation reaction can both {\it `damp'} and {\it anti}-%
{\it `damp' }particle motion, though the average effect is usually a damping
one. On the other hand, we show that the fluctuations around the
mean-trajectory are damped, as described by a FDR. At the quantum stochastic
level as discussed here, quantum dissipation effect results from the {\it %
changes} in the radiation reaction force that are associated with
fluctuations in the particle trajectory around the mean, and therefore it is 
{\it not }the same force as the classical (i.e. average) radiation reaction
force. Only these quantum processes obey a FDR. This is a subtle noteworthy
point.

\subsection{Prior work in relation to ours}

\begin{enumerate}
\item  There are many works on {\it non}relativistic quantum (and
semiclassical) radiation reaction (for atoms as well as charges moving in
quantum fields) including those by Rohrlich, Moniz-Sharp, Cohen-Tannoudji
et. al., Milonni, and others \cite
{MilonniQVacuum94,Rohrlich69(Rad.Reaction.Charged.Particle),MonizSharp77,Cohen-Tannoudji92(Atom-Photon-Interactions),Milonni81}%
. Kampen, Moniz, Sharp, and others \cite{ExtendedCharges} suggested that the
problems of causality and runaways can be resolved in both classical and
quantum theory by considering extended charge models. However, this
approach, while quite interesting, misses the point that point particles
(like local quantum field theories) obey a good low energy effective theory
in their own right. While it is true that extended objects can cure
infinities (e.g., extended charge models, string theory), it is nonetheless
important to recognize that at sufficiently low energy QED as an effective
field theory is `consistent' without recourse to its high-energy limit.
Wilson, Weinberg and others \cite{EFT} have shown how effective theory
description is sufficient to understand low energy physics because
complicated, and often irrelevant, high energy details of the fine structure
are not being probed at the physical energy scales of interest. The
effective theory approach has also provided a new perspective on the
physical meaning of renormalization and the source of divergences, showing
why nonrenormalizable interactions {\it are not }a disaster, why the
apparent large shift in bare-parameters resulting from divergent loop
diagrams (i.e. renormalization) doesn't invalidate perturbation theory, and
clarifying how and when high-energy structure {\it does }effect low-energy
physics. Therefore one should not need to invoke extended charge theory to
understand low energy particle dynamics.

Our work is a relativistic treatment which includes causal Non-Markovian
behavior and self-consistent backreaction. We think this is a better
approach even in a nonrelativistic context because the regularized
relativistic theory never `breaks down'. In our approach we adopt the ideas
and methods of effective field theory with proper treatment of backreaction
to account for the effects of high-energy (short-distance) structure on
low-energy behavior, and to demonstrate the self-consistency of the
semiclassical particle dynamics. We consider the consequences of
coarse-graining the irrelevant (environmental) degrees of freedom, and the
nature of vacuum fluctuations and quantum dissipation in the radiation
reaction problem. We emphasize the crucial role of decoherence due to noise
in resolving the pathologies and paradoxes, and in seeing how causal QED
leads to a (short-time modified) causal ALD semiclassical limit.

\item  By examining the time-dependence of operator canonical commutation
relations Milonni showed the necessity of electromagnetic field vacuum
fluctuations for radiating {\it nonrelativistic} charges \cite{Milonni81}.
The conservation of the canonical commutation relations is a fundamental
requirement of the quantum theory. Yet, if a quantum particle is coupled to
a {\it classical} electromagnetic field, radiative losses (dissipation) lead
to a contraction in ``phase-space'' for the expected values of the particle
position and momentum, which violates the commutation relations. It is the
vacuum field that balances the dissipation effect, preserving the
commutation relations $\left[ {\bf \hat{x},\hat{p}}\right] =i\hbar$ as a
consequence of a fluctuation-dissipation relation (FDR). The
particle-centric worldline framework allows us to consider similar issues
for relativistic particles. Interestingly, our covariant framework shows how
quantum fluctuations (manifesting as noise) appear in both the time and
space coordinates of a particle. This is not surprising since relativity
requires that a physical particle satisfies $\dot{z}^{\mu}\dot{z}_{\mu}=\dot{%
t}^{2}\left( \tau\right) -{\bf \dot{z}}^{2}\left( \tau\right) =1,$ and
therefore any variation in spatial velocities must be balanced by a change
in $\dot{t}^{0}$ keeping the particle ``on-shell''.

\item  The derivation and use of quantum Langevin equations (QLE's) to
describe fluctuations of a system in contact with a quantum environment has
a long history. Typically, QLE's are assumed to describe fluctuations in the
linear response regime for a system around equilibrium, but its validity
does not need be so restrictive. Nonequilibrium conditions can be treated
with the Feynman-Vernon influence functional. Caldeira and Leggett's study
of quantum Brownian motion (QBM) has led to an extensive literature \cite
{QBM}, particularly in regard to decoherence issues\cite{EarlyDecoherence}.
Barone and Caldeira \cite{BaroneCaldeira} have applied this method to the
question of whether nonrelativistic, dipole coupled electrons decohere in a
quantum electromagnetic field. An advantage of Barone and Caldeira's work is
that it is not limited to initially factorized states; they use the
preparation function method which allows the inclusion of initial particle
field correlations. Despite this, Romero and Paz have pointed out that the
preparation function method still suffers from an (implicit) unphysical
depiction of instantaneous measurement characterizing the initial state
preparation \cite{RomeroPaz}.

\item  Ford, Lewis, and O'Connell have extensively discussed the
electromagnetic field as a thermal bath in the linear, dipole coupled regime 
\cite{FOL}, and pioneered the application of QLE's to nonrelativistic
particle motion in QED. They have detailed the conditions for causality in
the thermodynamic, equilibrium limit described by the late-time linear
quantum Langevin equation. A crucial point of their analysis is that
particle motion can be (depending on the cutoff of the field spectral
density) runaway free and causal in the late-time limit \cite{FO(runaways)}.
In \cite{FO(FDR)}, they suggest a form of the equations of motion that gives
fluctuations without dissipation for a free electron, but this result is
special to the particular choice of a field cutoff made by the authors,
namely, one implying that the bare mass of the particle exactly vanishes.
When one does not assume a special value for the cutoff one generally does
find both fluctuations and dissipation, though Ford and Lewis's
counter-example shows how careful one must be in handling FDRs. Another
counter-example against making hasty generalizations is the vanishing of the
averaged (classical) radiation reaction for uniformly accelerated particles
despite quantum fluctuations that make the (suitably coarse-grained)
trajectories stochastic.

We do agree with the result that the particle motion is only causal and
consistent when the field cutoff is below a certain critical value\footnote{%
A noted example is in Caldeira and Leggett's original work \cite
{CaldeiraLeggett}. Their master equation governs a reduced density matrix
which is not positive definite because they choose an (infinite) cutoff
greater than the inverse temperature (see Hu, Paz, and Zhang \cite{QBM} for
details).} (determined from the particles classical radius), but we find no
compelling justification for the claim that the cutoff {\it should }take
exactly that special value.

It is certainly interesting that with the special field-cutoff chosen by
Ford and O'Connell higher time-derivatives vanish from the equations of
motion. However, even without such a specially chosen cutoff, the influence
of higher derivative terms are strongly suppressed at low energies. On this
issue we take the effective theory point of view which emphasizes the
generic insensitivity of low-energy phenomena to unobserved high-energy
structures. Since higher-time derivatives in the equations of motion do not
(necessarily) violate causality, there is no reason to rule them out. It is
perhaps instructive to review the meaning of renormalizable field theories
in the light of effective theories. Effective theories generically have
higher derivative interactions that are again strongly suppressed at low
energies. What makes renormalizable terms special is that these are the
interactions that remain relevant (or marginal) at low energies, while
non-renormalizable terms (such as higher derivative terms) are exponentially
suppressed, hence, it is no longer believed that non-renormalizable terms
are fundamentally excluded so long as one recognizes that physical theories
are mostly effective theories. Since QED is certainly an effective theory,
these observations (together with the suppression of the higher
time-derivative terms) weaken the argument that the special cutoff theory of
FOL is fundamental. Ford and O'Connell also propose a relativistic
generalization of their modified equations of motion for the average
trajectory derived from the nonrelativistic QLE \cite{FO(relativistic)}. Our
derivation of the stochastic limit from a relativistic quantum mechanics
(i.e. worldline formulation) and field theory goes well beyond this by
allowing the treatment of fully quantum relativistic processes, and by
deriving a relativistic Langevin equation.

\item  A perturbative expansion of the ALD equation up to order $e^{3}$ has
been derived from QED field theory by Krivitskii and Tsytovich \cite
{Krivistskii.Tsytovich91(Average.RR.QED)}, including the additional forces
arising from particle spin. Their work shows that the ALD equation may also
be understood from field theory, but the authors have not addressed the role
of fluctuations, correlation, decoherence, time-dependent renormalization,
nor self-consistent backreaction. Our derivation yields the full ALD
equation (all orders in $e$ for the perturbation expansion employed by \cite
{Krivistskii.Tsytovich91(Average.RR.QED)}) through the loop (semiclassical)
expansion; this method automatically includes all tree-level diagrammatic
effects.

\item  Low \cite{Low98:runaways} showed that runaway solutions apparently do
not occur in spin 1/2 QED. But Low does not derive the ALD\ equation, nor
address the semiclassical/stochastic limit. We also emphasize, again, that
it does not, and should not, matter whether particles are spin 1/2, spin 0,
or have some other internal structure in regard to the causality of the low
energy effective theory for center of mass particle motion.

\item  Using the influence functional, Di\'{o}si \cite{Diosi} derives a
Markovian master equation in non-relativistic quantum mechanics. In
contrast, it is our intent to emphasize the non-Markovian and nonequilibrium
regimes with special attention paid to self-consistency. The work of \cite
{Diosi} differs from ours in the treatment of the influence functional as a
functional of particle trajectories in the relativistic worldline
quantization framework. Ford has considered the loss of electron coherence
from vacuum fluctuation induced noise with the same noise kernel that we
employ \cite{Ford}. However, his application concerns the case of fixed or
predetermined trajectories.
\end{enumerate}

\subsection{Organization}

In Section II we obtain the influence functional (IF) for spinless
relativistic particles. We then define the stochastic effective action $%
\left( S_{stoch}\right) $ for this model, using our results from Paper I,
and use it to derive nonlinear integral Langevin equations for multiparticle
spacetime motion. In Section III we consider the single particle case. We
derive the (scalar-field) Abraham-Lorentz-Dirac (ALD) equation as the
self-consistent semiclassical limit, and show how this limit emerges free of
all pathologies. In Section IV we derive a Langevin equation for the
stochastic fluctuations of the particle spacetime coordinates about the
semiclassical limit for both the one particle and multiparticle cases. We
also show how a stochastic version of the Ward-Takahashi identities
preserves the mass-shell constraint. In Section V, we give a simple example
of these equations for a single free particle in a scalar field. We find
that in this particular case the quantum field induced noise vanishes,
though this result is special to the overly-simple scalar field, and does
not hold for the electromagnetic field \cite{JH3}. In the final section, we
summarize our main results and mention areas of applications, of both
theoretical and practical interest.

\section{Spinless relativistic particles moving in a scalar field}

\subsection{The model}

Relativistic quantum theories are usually focused primarily on quantized
fields, the notion of particles following trajectories is somewhat
secondary. As explained in Paper I \cite{JH1}, we employ a ``hybrid'' model
in which the environment is a field, but the system is the collection of
particle spacetime coordinates $z_{n}^{\mu}\left( \tau_{n}\right) $ (i.e.,
worldlines) where $n$ indicates the $n^{th}$ particle coordinate, with
worldline parameter $\tau_{n},$ charge $e_{n},$ and mass $m_{n}.$ The free
particle action is 
\begin{equation}
S_{A}[z]=\sum_{n}\int d\tau_{n}\sqrt{\left( \dot{z}_{n}^{\mu}\right) ^{2}}%
\left[ m_{n}+V(z_{n}(\tau_{n}))\right] .  \label{Square-root action}
\end{equation}
From $S_{A}$ follow the relativistic equations of motion: 
\begin{equation}
m\dot{z}_{n}^{\mu}/\sqrt{\dot{z}_{n}^{2}}=-\partial^{\mu}V\left(
z_{n}\right) .  \label{system equations of motion}
\end{equation}
For generality, we include a possible background potential $V,$ in addition
to the quantum scalar field environment. The scalar current is 
\begin{equation}
j[z,x)=\sum_{n}\int d\tau_{n}e_{n}u_{n}(\tau_{n})\delta(x-z_{n}(\tau _{n})),
\label{current}
\end{equation}
where 
\begin{equation}
u_{n}\left( \tau_{n}\right) =\sqrt{\dot{z}_{n}^{\mu}\left( \tau_{n}\right) 
\dot{z}_{n,\mu}\left( \tau_{n}\right) }.  \label{Definition of u}
\end{equation}
In Paper III \cite{JH3}, we treat a vector current coupled to the
electromagnetic vector potential $A_{\mu}$. In both cases we assume spinless
particles. The inclusion of spin or color is important to making full use of
these methods in QED and QCD.

In the full quantum theory reparametrization invariance of the relativistic
particle plays a crucial role, and the particle-field interaction must
respect this symmetry. When dealing with the path integral for the quantized
particle worldline, a quadratic form of the action is often more convenient
than the square-root action in (\ref{Square-root action}). The worldline
quantum theory is a gauge theory because of the constraint that follows from
reparametrization invariance under $\tau\rightarrow\tau^{\prime}\left(
\tau\right) =\tau+\varepsilon\left( \tau\right) ,$ making this perhaps the
simplest physical example of a generally covariant theory. (We may therefore
view this work as a toy model for general relativity, and for string theory,
in the semiclassical and stochastic regime.) The path integral must
therefore be gauge-fixed to prevent summing over gauge-equivalent histories.
In \cite{JH4-5} we treat the full quantum theory, in detail, and derive the
``worldline influence functional'' after fixing the gauge to the so-called
proper-time gauge. This give the worldline path integral in terms of a sum
over particle trajectories $z^{\mu}\left( \tau\right) $ satisfying $\dot {z}%
^{2}=N^{2},$ but where the path integral still involves an integration over
possible $N$ (the range of this integration depends on the boundary
conditions for the problem at hand). However, applying the semiclassical
(i.e., loop) expansion in \cite{JH4-5} we find that the $N=1$ trajectory
gives the stationary phase solution; hence, the semiclassical limit\footnote{%
Note however that in the stochastic limit there can potentially be
fluctuations in the value of the constraint.} corresponds to setting $N=1.$
In the following we will therefore assume the constraint $\dot{z}^{2}=N^{2},$
or even $\dot {z}^{2}=1,$ when we are discussing the semiclassical limit for
trajectories. We show that this is self-consistent in that it is preserved
by the equations of motion including noise-induced fluctuations in the
particle trajectory. Since we are addressing the semiclassical/stochastic
limit in this paper there is no need to adopt the quadratic form of action
for a relativistic particle; it is adequate, and it turns out to be simpler
in this paper, to just work with the scalar functions $u_{n}$ defined in (%
\ref{Definition of u}), and the square-root form of the relativistic
particle action in (\ref{Square-root action}).

While we emphasize the microscopic quantum origins, we might also view our
model in analogy to the treatment of quantum fields in curved spacetime.
There, one takes the gravitational field (spacetime) as a classical system
coupled to quantum fields. One important class of problems is then the
backreaction of quantum fields on the classical spacetime. The backreaction
of their mean yields the semiclassical Einstein equation which forms the
basis of semiclassical gravity. The inclusion of fluctuations of the
stress-energy of the quantum fields and the induced metric fluctuations
yields the Einstein-Langevin equations \cite{HM95EinsteinLangevin} which
forms the basis of this stochastic semiclassical gravity \cite
{Hu99StochasticGravity}. In our work here, the particle coordinates are
analogous to the gravitational field (metric tensor). Whereas the
Einstein-Langevin equations have not been derived from first principles for
the lack of a quantum theory of gravity, we shall take advantage of the
existence of a full quantum theory of particles and fields to examine how,
and when, a regime of stochastic behavior emerges.

\subsection{The influence functional and stochastic effective action}

The interaction between the particles and a scalar field is given by the
monopole coupling term 
\begin{align}
S_{int} & =\int dx\,j[z,x)\,\varphi(x)  \label{interaction action} \\
& =e\sum_{n}\int d\tau_{n}u_{n}(\tau_{n})\varphi(z_{n}(\tau_{n})).  \nonumber
\end{align}
This is the general type of interaction treated in Paper I. In the second
line we have used the expression for the current (\ref{current}). The
expression for the influence functional (Eq. (3.16) of Paper I) \cite
{FeynmanVernon} is, 
\begin{align}
F[j^{\pm}] & =\text{$\exp${\LARGE \{}}-\frac{i}{\hbar}\int dx\int dx^{\prime
}\text{{\large [}}2j^{-}(x)\,G^{R}(x,x^{\prime})j^{+}(x^{\prime})  \nonumber
\\
& -ij^{-}(x)G^{H}(x^{\prime},x^{\prime})j^{-}(x^{\prime})\text{{\large ]}%
{\LARGE \}}},  \label{F[j,j'] in terms of Green's functions}
\end{align}
where 
\begin{align}
j^{-} & \equiv\left( j[z,x)-j[z^{\prime},x)\right) \\
j^{+} & \equiv\left( j[z,x)+j[z^{\prime},x)\right) /2  \nonumber
\end{align}
and $\{z,$ $z^{\prime}\}$ are a pair of particle histories. $G^{R}$ and $%
G^{H}$ are the scalar field retarded and Hadamard Green's functions,
respectively. Substitution of (\ref{current}) then gives the multiparticle
influence functional 
\begin{align}
F[\left\{ z\right\} ,\left\{ z^{\prime}\right\} ] & =\text{$\exp ${\LARGE \{}%
}-\frac{e^{2}}{\hbar}\sum_{nm}\int d\tau_{n}\int d\tau _{m}\,
\label{F[z,z']: scalar particle IF 1} \\
& \times\theta(T-z_{n}^{0})\theta(z_{n}^{0}-z_{m}^{0})  \nonumber \\
& \times\text{{\Large [}}u_{n}G^{H}(z_{n},z_{m})u_{m}-u_{n}^{%
\prime}G^{H}(z_{n}^{\prime},z_{m})u_{m}  \nonumber \\
&
-u_{n}G^{H}(z_{n},z_{m}^{\prime})u_{m}^{\prime}+u_{n}^{\prime}G^{H}(z_{n}^{%
\prime},z_{m}^{\prime})u_{m}^{\prime}  \nonumber \\
&
+u_{n}G^{R}(z_{n},z_{m})u_{m}-u_{n}^{\prime}G^{R}(z_{n}^{\prime},z_{m})u_{m}
\nonumber \\
&
+u_{n}G^{R}(z_{n},z_{m}^{\prime})u_{m}^{\prime}-u_{n}^{\prime}G^{R}(z_{n}^{%
\prime},z_{m}^{\prime})u_{m}^{\prime}\text{{\Large ]}{\LARGE \}}}  \nonumber
\end{align}
where $u_{n}^{\prime}=u_{n}(z_{n}^{\prime}).$ The influence functional may
be expressed more compactly by using a matrix notation where ${\bf u}%
_{n}^{T}=\left( u_{n},u_{n}^{\prime}\right) \equiv\left(
u_{n}^{1},u_{n}^{2}\right) $, giving 
\begin{align}
F[z^{a}] & =\text{$\exp${\LARGE \{}}-\frac{e^{2}}{\hbar}\sum_{nm}\int
d\tau_{n}d\tau_{m}  \label{F[za,zb]: scalar particle IF 2} \\
& \times\left( {\bf u}_{n}^{T}{\bf G}_{nm}^{R}{\bf u}_{m}+{\bf u}_{n}^{T}%
{\bf G}_{nm}^{H}{\bf u}_{m}\right) \text{{\LARGE \}}}  \nonumber \\
& =\text{$\exp$}\left\{ -\frac{e^{2}}{\hbar}\left( {\bf u}_{n}^{T}{\bf G}%
_{nm}^{R}{\bf u}_{m}+{\bf u}_{n}^{T}{\bf G}_{nm}^{H}{\bf u}_{m}\right)
\right\}  \label{Suppression of sum and integration} \\
& =\text{$\exp$}\left\{ \frac{i}{\hbar}S_{IF}[z^{a}]\right\} .
\label{particle influence action}
\end{align}
The superscript $T$ denotes the transpose of the column vector ${\bf u.}$ In
(\ref{Suppression of sum and integration}), and below, we leave the sum $%
\Sigma_{nm}$ and integrations $\int d\tau_{n}d\tau_{m}$ implicit for
brevity. The matrices {\bf $G$}$_{nm}^{H},${\bf $G$}$_{nm}^{R}$ are given by 
\begin{align}
{\bf G}_{nm}^{H} & =\theta(T-z_{n}^{0})\theta(z_{n}^{0}-z_{m}^{0})
\label{Gn(ab): scalar particle noise matrix} \\
& \times\left( 
\begin{array}{cc}
G_{(11)}^{H}\left( z_{n}^{1},z_{m}^{1}\right) & -G_{(12)}^{H}\left(
z_{n}^{1},z_{m}^{2}\right) \\ 
-G_{(21)}^{H}\left( z_{n}^{2},z_{m}^{1}\right) & G_{(22)}^{H}\left(
z_{n}^{2},z_{m}^{2}\right)
\end{array}
\right)  \nonumber
\end{align}
and 
\begin{align}
{\bf G}_{nm}^{R} & =\theta(T-z_{n}^{0})\theta(z_{n}^{0}-z_{m}^{0})
\label{Gr(zb): scalar particle dissipation matrix} \\
& \times\left( 
\begin{array}{cc}
G_{(11)}^{R}\left( z_{n}^{1},z_{m}^{1}\right) & G_{(12)}^{R}\left(
z_{n}^{1},z_{m}^{2}\right) \\ 
-G_{(21)}^{R}\left( z_{n}^{2},z_{m}^{1}\right) & -G_{(22)}^{R}\left(
z_{n}^{2},z_{m}^{2}\right)
\end{array}
\right) .  \nonumber
\end{align}
In Eq. (\ref{particle influence action}), we have defined the influence
action $S_{IF}$. $G^{H,R}$ are the scalar-field Hadamard/Retarded Green's
functions evaluated at various combinations of spacetime points $%
z_{n,m}^{1,2}.$

The stochastic effective action is defined by (see \cite{JH1}) 
\begin{align}
S_{\chi}[z^{\pm}] & =S_{A}[z^{\pm}]+\int dxj^{-}[z^{\pm},x)(\chi(x) \\
& +2\int dx^{\prime}G^{R}(x,x^{\prime})j^{+}[z^{\pm},x^{\prime}))  \nonumber
\\
& =S_{A}[z^{a}]+{\bf u}_{n}^{T}{\bf G}_{nm}^{R}{\bf u}_{m}+{\bf u}_{n}^{T}%
{\bf \varpi}_{n},  \nonumber
\end{align}
where 
\begin{equation}
{\bf u}_{n}^{T}{\bf \varpi}_{n}=u_{n}\left( z_{n}\right) \chi\left(
z_{n}\right) -u_{n}\left( z_{n}^{\prime}\right) \chi\left( z_{n}^{\prime
}\right) ,
\end{equation}
and $\chi\left( z_{n}\right) $ is a (classical) stochastic field (see
Section 3 of Paper I) evaluated at the spacetime position of the $n$th
particle. The stochastic field has vanishing mean and autocorrelation
function given by 
\begin{equation}
\langle\chi\left( x\right) \chi\left( x^{\prime}\right) \rangle
_{s}=\hslash\langle\left\{ \hat{\varphi}\left( x\right) ,\hat{\varphi }%
\left( x^{\prime}\right) \right\} \rangle=\hslash G^{H}\left(
x,x^{\prime}\right) .
\end{equation}
Hence, $\chi\left( x\right) ^{\prime}s$ statistics encodes those of the
quantum field $\hat{\varphi}\left( x\right) $.

The stochastic effective action provides an efficient means for deriving
self-consistent Langevin equations that describe the effects of quantum
fluctuations as stochastic particle motion for sufficiently decohered
trajectories (for justification in using $S_{stoch}$ in this way see Paper
I).

\subsection{Langevin integrodifferential equations of motion}

Deriving a specific Langevin equation from the general formalism of Paper I
requires evaluating the cumulants $C_{q}^{\left( \eta\right) },$ defined in
Eq. (4.11) of \cite{JH1}. Because the multiparticle case involves a
straightforward generalization, we begin with the single particle theory.
The noise cumulants are then defined as 
\begin{equation}
C_{q}^{\left( \eta\right) }=\left( \frac{\hslash}{i}\right) ^{q}\left. \frac{%
\delta^{q}S_{IF}\left[ j^{\pm}\right] }{\delta z_{n}^{\mu-}\left(
\tau_{n}\right) ...\delta z_{m}^{\nu-}\left( \tau_{m}\right) }\right|
_{z^{-}=0},
\end{equation}
with $S_{IF}$ given in (\ref{particle influence action}). The superscript $%
\left( \eta\right) $ indicates that these are cumulants for an expansion of $%
S_{IF}$ with respect to the stochastic variables $\eta^{\mu}\left(
\tau\right) .$

We define new variables ${\bf v=Uu}$ where ${\bf v}^{T}{\bf =}\left(
v^{1},v^{2}\right) $ and ${\bf u}^{T}{\bf =}\left( u,u^{\prime }\right) ,$
and 
\begin{equation}
{\bf U=}\frac{1}{2}\left( 
\begin{array}{cc}
1 & -1 \\ 
1 & 1
\end{array}
\right) .
\end{equation}
Then the influence action has the form 
\begin{equation}
S_{IF}[z^{a}]=\frac{e^{2}}{\hslash}\left( {\bf v}^{T}{\bf G}_{v}^{R}{\bf v}+i%
{\bf v}^{T}{\bf G}_{v}^{H}{\bf v}\right) ,
\label{Influence function in matrix form}
\end{equation}
where 
\begin{align}
{\bf G}_{v}^{R} & ={\bf UG}^{R}{\bf U}^{T}  \label{Gr matrix} \\
& =\left( 
\begin{array}{cc}
{\small G}_{{\small 11}}^{{\small R}}{\small -G}_{{\small 12}}^{{\small R}}%
{\small +G}_{{\small 21}}^{{\small R}}{\small -G}_{{\small 22}}^{{\small R}}
& {\small G}_{{\small 11}}^{{\small R}}{\small +G}_{{\small 12}}^{{\small R}}%
{\small +G}_{{\small 21}}^{{\small R}}{\small +G}_{{\small 22}}^{{\small R}}
\\ 
{\small G}_{{\small 11}}^{{\small R}}{\small -G}_{{\small 12}}^{{\small R}}%
{\small -G}_{{\small 21}}^{{\small R}}{\small +G}_{{\small 22}}^{{\small R}}
& {\small G}_{{\small 11}}^{{\small R}}{\small +G}_{{\small 12}}^{{\small R}}%
{\small -G}_{{\small 21}}^{{\small R}}{\small -G}_{{\small 22}}^{{\small R}}
\end{array}
\right) ,  \nonumber
\end{align}
and 
\begin{align}
{\bf G}_{v}^{H} & ={\bf UG}^{H}{\bf U}^{T}  \label{Gh matrix 1} \\
& =\left( 
\begin{array}{cc}
{\small G}_{{\small 11}}^{{\small H}}{\small +G}_{{\small 12}}^{{\small H}}%
{\small +G}_{{\small 21}}^{{\small H}}{\small +G}_{{\small 22}}^{{\small H}}
& {\small G}_{{\small 11}}^{{\small H}}{\small -G}_{{\small 12}}^{{\small H}}%
{\small +G}_{{\small 21}}^{{\small H}}{\small -G}_{{\small 22}}^{{\small H}}
\\ 
{\small G}_{{\small 11}}^{{\small H}}{\small +G}_{{\small 12}}^{{\small H}}%
{\small -G}_{{\small 21}}^{{\small H}}{\small -G}_{{\small 22}}^{{\small H}}
& {\small G}_{{\small 11}}^{{\small H}}{\small -G}_{{\small 12}}^{{\small H}}%
{\small -G}_{{\small 21}}^{{\small H}}{\small +G}_{{\small 22}}^{{\small H}}
\end{array}
\right)  \nonumber
\end{align}

The lowest order cumulant in the Langevin equation, $C_{1,\mu}^{\left(
\eta\right) },$ is found by evaluating $\delta S_{IF}/\delta z^{\mu-}$, and
then setting $z^{-}=0.$ There are two kinds of terms that arise: those where 
$\delta/\delta z^{\mu-}$ acts on ${\bf v,}$ and those where it acts on {\bf $%
G$}$^{R,H}.$ For linear theories, {\bf $G$}$^{R,H}$ is not a function of the
dynamical variables, but is instead (at most) a function of some
predetermined kinematical variables that a priori specify the trajectory.
Hence, there is no contribution from $\delta${\bf $G$}$^{R,H}/\delta
z^{\mu-} $. For the $\delta{\bf v}/\delta z^{\mu-}$ terms, setting $z^{-}=0$
collapses the matrices (\ref{Gr matrix} and \ref{Gh matrix 1}) to one term
each: only the $\left( 1,2\right) $ term of {\bf $G$}$_{v}^{R},$ and the $%
\left( 1,1\right) $ term of {\bf $G$}$_{v}^{R},$ survive. But, setting $%
z^{-}=0$ gives $v^{1}=0,$ $v^{2}=u,$ and since the $\left( 1,1\right) $ term
of {\bf $G$}$_{v}^{H}$ is proportional to two factors of $v^{1},$ it also
vanishes. When $\delta/\delta z^{\mu-}$ acts on {\bf $G$}$_{v}^{R},$ only
the $\left( 2,2\right) $ terms survive (because it is the only term not
proportional to a factor of $v^{1}).$ For similar reasons, the only
contributing element of $G^{H}$ is also its $\left( 2,2\right) $ term. To
evaluate $\left( \delta G_{v\left( 22\right) }^{R}/\delta z^{\mu-}\right)
|_{z^{-}=0},$ we note 
\begin{equation}
\frac{\delta}{\delta z^{-\mu}}=\frac{1}{2}\left\{ \frac{\delta}{\delta z}-%
\frac{\delta}{\delta z^{\mu\prime}}\right\} .
\end{equation}
After a little algebra, we are left with 
\begin{equation}
\left( \delta G_{v\left( 22\right) }^{R}/\delta z^{\mu-}\right) |_{z^{-}=0}=%
\frac{\delta G^{R}(z(\tau),z(\tau^{\prime}))}{\delta z^{\mu}(\tau)},
\end{equation}
where the derivative only acts on the $z(\tau),$ and not the $z(\tau^{\prime
})$ argument, in $G^{R}.$ The same algebra in evaluating the $\left( \delta
G_{v\left( 22\right) }^{H}/\delta z^{\mu-}\right) |_{z^{-}=0}$ term shows
that all the factors cancel, and therefore $G^{H}$ does not contribute to
the first cumulant at all. Because the imaginary part of $S_{IF}[z^{a}]$
does not contribute to the first cumulant, the equations of motion of the
mean-trajectory are explicitly real, which is an important consequence of
using an initial value formulation like the influence functional (or
closed-time-path) method.

The first cumulant, describing {\it radiation reaction}, is therefore given
by 
\begin{align}
& C_{1\mu}^{\left( \eta\right) }[z,\tau)=\left. \frac{\delta S_{IF}}{\delta
z^{\mu-}(\tau)}\right| _{z^{-}=0} \\
& =\int_{\tau_{i}}^{\tau}d\tau^{\prime}e^{2}\text{{\LARGE \{}}\left( \frac{%
\delta u(z)}{\delta z^{\mu}\left( \tau\right) }\right)
G^{R}(z(\tau),z(\tau^{\prime}))u(z(\tau^{\prime}))  \nonumber \\
& +u(z(\tau))\left( \frac{\delta G^{R}(z(\tau),z(\tau^{\prime}))}{\delta
z^{\mu}(\tau)}\right) u(z(\tau^{\prime}))\text{{\LARGE \}}}{\Huge .} 
\nonumber
\end{align}
This expression is explicitly causal both because the proper-time
integration is only over values $\tau^{\prime}<\tau$, and because of the
explicit occurrence of the retarded Green's function. Contrary to common
perception, the radiation reaction force given by $C_{1\mu}^{\left(
\eta\right) }$ is not necessarily dissipative in nature. For instance, we
shall see that $C_{1\mu}^{\left( \eta\right) }$ vanishes for uniformly
accelerated motion, despite the presence of radiation from a uniformly
accelerating charge. In other circumstances, $C_{1\mu}^{\left( \eta\right)
}[z,\tau)$ may actually provide an anti-damping force for some portions of
the particle trajectory.

Hence, radiation reaction is not a purely dissipative, or damping, process.
The relationship of radiation reaction to vacuum fluctuations is also not
simply given by a fluctuation-dissipation relation\footnote{%
Fluctuation-dissipation related arguments have been frequently invoked for
radiation reaction problems, but these instances generally entail
linearizing (say, making the dipole approximation), and/or the presence of a
binding potential that provide a restoring force to the charge motion making
it periodic (or quasi-periodic). Radiation reaction in atomic systems is one
example of this \cite{Cohen-Tannoudji92(Atom-Photon-Interactions)}; the
assumption of some kind of harmonic binding potential provides another
example \cite{Milonni81,CallenWelton51(Fluct.Diss.Relations)}.}. In
actuality, only that part of the first cumulant describing {\it deviations}
from the purely classical radiation reaction force is balanced by
fluctuations in the quantum field via generalized fluctuation dissipation
relations. This part we term {\it quantum dissipation} because it is of
quantum origin and different from classical radiation reaction. Making this
distinction clear is important to dispel common misconceptions about
radiation reaction being always balanced by vacuum fluctuations.

Next we evaluate the second cumulant. After similar manipulations as above,
we find that the second cumulant involves only $G^{H}.$ We note that the
action of $\delta j/\delta z^{\mu}$ on an arbitrary function is given by 
\begin{align}
& \int dx\frac{\delta j(x)}{\delta z^{\mu}(\tau)}f(x) \\
& =e\int dx\int d\tau\frac{\delta}{\delta z^{\mu}}\left( \left( \dot{z}%
^{2}\right) ^{1/2}\delta(x-z(\tau))\right) f(x)  \nonumber \\
& =e\text{{\LARGE \{}}\frac{d}{d\tau}\left( \frac{\dot{z}_{\mu}}{u(\tau )}%
\right) +\left( \frac{\dot{z}_{\mu}}{u(\tau)}\right) \frac{d}{d\tau }-u(\tau)%
\frac{\partial}{\partial z^{\mu}}\text{{\LARGE \}}}f(z(\tau ))  \nonumber
\end{align}
Because of the constraint $\dot{z}^{2}=N^{2},$ it follows that $\dot{u}=0,$
and we may set $u=N.$ Also, $\dot{z}^{\mu}\ddot{z}_{\mu}=0.$ Then 
\begin{align}
\int dx\frac{\delta j(x)}{\delta z^{\mu}(\tau)}f(x) & =e\left[ \frac{1}{N}%
\left\{ \ddot{z}_{\mu}+\dot{z}^{\nu}\dot{z}_{[\mu}\partial_{\nu]}\right\} %
\right] f(z(\tau))  \nonumber \\
& \equiv e\vec{w}_{\mu}\left( z\right) f(z(\tau)).
\end{align}
This last expression defines the operator $\vec{w}_{\mu}\left( z\right) .$
We have used $d/d\tau=\dot{z}_{\nu}\partial^{\nu}.$ The operator $\vec{w}%
_{\mu}$ satisfies the identity 
\begin{align}
\dot{z}^{\mu}\vec{w}_{\mu}\left( z\right) & =\dot{z}^{\mu}\left[ \frac {1}{N}%
\left\{ \ddot{z}_{\mu}+\dot{z}^{\nu}\dot{z}_{[\mu}\partial_{\nu ]}\right\} %
\right]  \nonumber \\
& =\frac{1}{N}\left( \dot{z}^{\mu}\ddot{z}_{\mu}+\dot{z}^{\mu}\dot{z}^{\nu }%
\dot{z}_{[\mu}\partial_{\nu]}\right) =0.  \label{z dot w identity}
\end{align}
This identity ensures that neither radiation reaction nor noise-induced
fluctuations in the particle's trajectory move the particle off-shell (i.e.,
the stochastic equations of motion preserve the constraint $\dot{z}%
^{2}=N^{2} $ for any constant $N$).

With these definitions, the noise $\eta^{\mu}(\tau)$ is defined by 
\begin{align}
\eta_{\mu}(\tau) & =e\hbar^{1/2}\vec{w}_{\mu}(z)\chi(z(\tau))  \nonumber \\
& =e\hbar^{1/2}\left[ \frac{1}{N}\left\{ \ddot{z}_{\mu}+\dot{z}^{\nu}\dot {z}%
_{[\mu}\partial_{\nu]}\right\} \right] \chi (z)
\label{definition of noise w chi 2}
\end{align}
and the second-order noise correlator by

\begin{align}
C_{2}^{\left( \eta\right) \mu\nu}[z;\tau,\tau^{\prime}) & =\left\langle
\left\{ \eta^{\mu}(\tau),\eta^{\nu}(\tau^{\prime})\right\} \right\rangle \\
& =e^{2}\hbar\vec{w}^{\mu}(z)\vec{w}^{\nu}(z^{\prime})\langle\left\{
\chi\left( z\left( \tau\right) \right) ,\chi\left( z\left( \tau^{\prime
}\right) \right) \right\} \rangle  \nonumber \\
& =e^{2}\hbar\vec{w}^{\mu}(z)\vec{w}^{\nu}(z^{\prime})G^{H}(z(\tau
),z(\tau^{\prime})).  \nonumber
\end{align}
The operator $\vec{w}^{\mu}\left( z\right) $ acts only on the $z$ in $%
G^{H}\left( z,z^{\prime}\right) ;$ likewise, the operator $\vec{w}^{\nu
}\left( z^{\prime}\right) $ acts only on $z^{\prime}.$

This scalar field result is reminiscent of electromagnetism, where the
Lorentz force from the (antisymmetric) field strength tensor $F_{\mu\nu}$ is 
${\cal F}_{\mu}$ $=\dot{z}^{\nu}F_{\mu\nu}=\dot{z}^{\nu}\partial
_{\lbrack\mu}A_{\nu]}$. The antisymmetry of $F_{\mu\nu}$ implies $\dot{z}%
^{\mu}{\cal F}_{\mu}=0$. We may define a scalar analog of the antisymmetric
(second rank) field strength tensor by 
\begin{equation}
F_{\mu\nu}^{\chi}\equiv\dot{z}_{[\mu}\partial_{\nu]}\chi.
\end{equation}
This shows that the second term on the right hand side of (\ref{definition
of noise w chi 2}) gives the scalar analog of (a stochastic) electromagnetic
Lorentz force: ${\cal F}_{\mu}^{\chi}=\dot{z}^{\nu}F_{\mu\nu}^{\chi}.$ The
first term on the RHS\ of (\ref{definition of noise w chi 2}), $\ddot{z}%
_{\mu }\chi\left( z\right) ,$ does not occur in the treatment of the
electromagnetic field. In the scalar-field theory, this term may be thought
of as a stochastic component to the particle mass.

The stochastic equations of motion are 
\begin{align}
m\ddot{z}_{\mu} & =-\partial_{\mu}V(z)+e^{2}\int^{\tau}d\tau^{\prime}\vec {w}%
_{\mu}(z)G^{R}(z(\tau),z(\tau^{\prime}))  \nonumber \\
& +e\hbar^{1/2}w_{\mu}(z)\chi(z(\tau)).  \label{Single particle Langevin 1}
\end{align}
The result (\ref{Single particle Langevin 1}) is formally a set of nonlinear
stochastic integrodifferential equations for the particle trajectories $%
z_{\mu}[\eta;\tau).$ Noise is absent in the classical limit found by the
prescription $\hbar\rightarrow0$\ (this definition of classicality is formal
in that the true semiclassical/classical limit requires coarse-graining and
decoherence, and is not just a matter of taking the limit $%
\hslash\rightarrow0).$

The generalization of (\ref{Single particle Langevin 1}) to multiparticles
is now straightforward. If we re-insert the particle number indices, the
first cumulant is 
\begin{align}
C_{1(n)\mu}^{\eta}[z,\tau) & =\left. \frac{\delta S_{IF}[z]}{\delta
z_{n}^{\mu-}(\tau)}\right| _{z_{m}^{-}=0} \\
& =\sum_{m}\int d\tau_{m}^{\prime}e^{2}\vec{w}_{n\mu}(z_{n})  \nonumber \\
& \times G^{R}(z_{n}(\tau_{n}),z_{m}(\tau_{m}^{\prime}))u(z_{m}(\tau
_{m}^{\prime}),  \nonumber
\end{align}
the noise term is given by 
\begin{equation}
\eta_{n}^{\mu}(\tau)=e\hbar^{1/2}w_{n}^{\mu}(z_{n})\vec{\chi}(z_{n}(\tau
_{n})),
\end{equation}
and the noise correlator is 
\begin{align}
& C_{2(nm)}^{\eta\mu\nu}[z;\tau_{n},\tau_{m}^{\prime}) \\
& =\left\langle \left\{ \eta_{n}^{\mu}(\tau_{n}),\eta_{m}^{\nu}(\tau
_{m}^{\prime})\right\} \right\rangle  \nonumber \\
& =e^{2}\hbar\vec{w}_{n}^{\mu}(z_{n})\vec{w}_{m}^{\nu}(z_{m})G^{H}(z_{n}(%
\tau_{n}),z_{m}(\tau_{m}^{\prime})).  \nonumber
\end{align}
The nonlinear multiparticle Langevin equations are therefore 
\begin{align}
& m\ddot{z}_{n\mu}\left( \tau\right)
\label{nonlinear multiparticle langevin equation} \\
& =-\partial_{\mu}V(z_{n}\left( \tau\right) )+e\hbar^{1/2}\vec{w}_{\mu
}(z_{n})\chi(z_{n}(\tau))  \nonumber \\
& +e^{2}\sum_{m}\int_{\tau_{i}}^{\tau_{f}}d\tau^{\prime}\vec{w}%
_{\mu}(z_{n}\left( \tau\right) )G^{R}(z_{n}(\tau),z_{m}(\tau^{\prime})). 
\nonumber
\end{align}

The $n\neq m$ terms in (\ref{nonlinear multiparticle langevin equation}) are
particle-particle interaction terms. Because of the appearance of the
retarded Green's function, all of these interactions are causal. The $n=m$
terms are the self-interaction (radiation-reaction) forces. The $n\neq m$
noise correlator terms represent nonlocal particle-particle correlations:
the noise that one particle sees is correlated with the noise that every
other particle sees. The nonlocally correlated stochastic field $\chi(x)$
reflects the correlated nature of the quantum vacuum. From the
fluctuation-dissipation relations found in Paper I (Eq. (5.23) and Eq.
(5.24)), the $n=m$ quantum noise is related to the $n=m$ quantum dissipative
forces. Under some, but {\it not }all, circumstances\footnote{%
See \cite{RavalHu:StochasticAcceleratedDetectors} and Paper I, Section IV,
for a discussion of this point. In brief, the noise correlation between
spacelike separated charges does not vanish owing to the nonlocality of
quantum theory, but the causal force terms involving $G^{R}$ does always
vanish between spacelike seperated points. These two kinds of terms are only
connected through a propagation/correlation relation (the multiparticle
generalization of an FDR) when one particle is in the other's casual future
(or past).} the $n\neq m$ correlation terms are likewise related to the $%
n\neq m$ propagation (interaction) terms through a multiparticle
generalization of the FDR, called a propagation-correlation relation \cite
{RavalHu:StochasticAcceleratedDetectors}.

We have already noted that the first term in (\ref{definition of noise w chi
2}) has the form of a stochastic contribution to the particle's effective
mass, thus allowing us to define the stochastic mass as 
\begin{equation}
m_{\chi}\,\ddot{z}^{\mu}=\left( m+e\hbar^{1/2}\chi(z)\right) \ddot{z}^{\mu}.
\end{equation}
Fluctuations of the stochastic mass automatically preserve the mass-shell
condition since $\left( m_{\chi}\ddot{z}^{\mu}\dot{z}_{\mu}\right) |_{z=\bar{%
z}}=0.$ Likewise, the effective stochastic force $F_{\mu}^{\chi}$ satisfies 
\begin{equation}
\dot{z}_{\mu}F_{\mu}^{\chi}=\dot{z}_{\mu}\dot{z}_{\nu}\dot{z}%
^{[\nu}\partial^{\mu]}\chi(z)=\dot{z}_{(\mu}\dot{z}_{\nu)}\dot{z}%
^{[\nu}\partial ^{\mu]}\chi(z)=0.
\end{equation}
These results show that the stochastic fluctuation-forces preserve the
constraint $\dot{z}^{2}=N^{2}.$

\section{The semiclassical regime: the scalar-field ALD equation}

We have emphasized in Paper I that the emergence of a Langevin equation (\ref
{Single particle Langevin 1} or \ref{nonlinear multiparticle langevin
equation}) presupposes decoherence, which works to suppress large
fluctuations away from the mean-trajectories. In the semiclassical limit,
the noise-average vanishes, and the semiclassical equations of motion for
the single particle are given by the mean of (\ref{Single particle Langevin
1}). The stochastic regime admits noise induced by quantum fluctuations
around these decohered mean solutions. In our second series of papers we
explore the stochastic behavior due to higher-order quantum effects that
refurbish the particle's quantum nature.

On a conceptual level, we note that the full quantum theory in the path
integral formulation involves summing over all worldlines of the particle%
\footnote{%
Actually, just what kinds of histories/worldlines are allowed in the sum is
determined by both the gauge choice in the worldline quantization method,
and by the boundary conditions.} joining the initial and final spacetime
positions, $z_{i}$ and $z_{f},$ respectively. Furthermore, the generating
functional for the worldline-coordinate expectation values (see Subsection
3.A of Paper I \cite{JH1}) involve a sum over final particle positions ${\bf %
z}_{f}.$ In these path integrals, there is no distinction between, say,
runaway trajectories and any other type of trajectory\footnote{%
In fact, trajectories in the path integral can be even stranger than the
runaways that appear in classical theory. Most paths are non-differentiable
(infinitely rough) and may even include those that travel outside the
lightcone and backward-in-time (see previous footnote). Clearly,
non-uniqueness of paths is not an issue in the path integral context.}.
Furthermore, no meaningful sense of causality is associated with individual
(i.e. fine-grained or skeletonized) histories in the path integral sum. Any
particular path in the sum going through the intermediate point $z\left(
\tau\right) $ at worldline parameter time $\tau$ bears no causal relation to
it than going through the point $z^{\prime}(\tau^{\prime})$ at some later
parameter time $\tau^{\prime}.$

Where questions of causality, uniqueness, and runaways do arise is in regard
to the solutions to the equations of motion for correlation function of the
worldline-coordinates, particularly for the expectation value giving the
mean trajectory. It is here that an initial value formulation for quantum
physics is crucial because only then are the equations of motion guaranteed
to be real and causal. In contrast, equations of motion found from the
in-out effective action (a transition amplitude formulation) are generally
neither real nor casual \cite
{Jordan86(EffectiveFieldEquationsExpectationValues,CalzettaHu87(CTP/backrection)}%
. Moreover, the equations of motion for correlation functions must be unique
and fully determined by the initial state if the theory is complete.

These observations lead to a reframing of the questions that are appropriate
in addressing the semiclassical limit of quantum particle-field
interactions. Namely, what are the salient features of the decoherent,
coarse-grained histories which qualify as semiclassical particle motion?
Quantum theory permits the occurrence of events which would be considered
classically improbable or forbidden in the particle motions (e.g., runaways,
tunneling, or apparently acausal behavior). These effects are allowed
because of quantum fluctuations. Bearing in mind the requirement of
classicality, we need to show that out of the infinite possibilities in the
quantum domain the interaction between particle and field (including the
self-field of the particle) is causal in the observable semiclassical limit.

So the pertinent questions in a quantum to stochastic treatment are the
following: What are the equations of motion for the mean and higher-order
correlation functions? Are {\it these }equations of motion causal and
well-defined? How significant are the quantum fluctuations around the
quantum-averaged trajectory? When does decoherence suppress the probability
of observing large fluctuations in the motion? When do the quantum
fluctuations assume a classical stochastic behavior? It is only by
addressing these questions that the true semiclassical motion may be
identified, together with the noise associated with quantum fluctuations
which is instrumental for decoherence. With this discussion as our guide, we
proceed in two steps. First, in this section, we find the semiclassical
limit for the equations of motion. Second, in the following section, we
describe the stochastic fluctuations around that limit. Since we are dealing
with a nonlinear theory, the fluctuations themselves must depend on the
semiclassical limit in a self-consistent fashion. These two steps constitute
the full backreaction problem for nonlinear particle field interactions in
the semiclassical and stochastic regimes. For more general conceptual
discussions on decoherent histories and semiclassical domains, see \cite
{GHDK,Decoherence,HLM95}.

\subsection{Divergences, regularized QED, and the initial state}

Even in classical field theory, Green's functions are usually divergent,
which calls for regularization. Ultraviolet divergences arise because the
field contains modes of arbitrarily short wavelengths to which a point
particle can couple. These facts have contributed to considerable confusion
about the correct form the radiation reaction force can take on for a
charged particle. There are also infrared divergences that arise from the
artificial distinction between soft and virtual photon emissions. These
divergences are an artifact of neglecting recoil on the particle motion.

In this work, we take the effective field theory philosophy as our guide.
One does not need to know the detailed structure of the correct high-energy
theory because low-energy processes are largely insensitive to them \cite
{EFT} beyond the effective renormalization of system parameters and
suppressed high-energy corrections. Take QED as an example, where relevant
physics occurs at the physical energy scale $E_{phys}$. In QED calculations,
it is generally necessary to assume a high-energy cutoff $E_{cutoff}$ that
regularizes the theory. One also assumes a high-energy scale $E_{eff}$ well
above $E_{phys}$ where presumably new physics is possible, thus $%
E_{phys}<<E_{eff}<E_{cutoff}.$ The effective theory that describes physics
at scales well below $E_{eff}$ is found by coarse-graining (integrating out)
the high-energy fluctuations between $E_{eff}$ and $E_{cutoff}.$ These
fluctuations renormalize the masses and charges in the effective theory, and
add new local, effective-interactions that reproduce the effects of the
coarse-grained physics between $E_{eff}$ and $E_{cutoff}.$ These so-called
nonrenormalizable effective interactions are suppressed by some power of $%
(E_{phys}/E_{eff}),$ making them usually insignificant compared to the
renormalizable terms in the effective theory. This justifies the viability
of ordinary QED, which is comprised of just those renormalizable terms. It
further ensures that physics based on the physical masses and charges probed
at $E_{phys}$ will be insensitive to modifications to QED above $E_{eff}$.

Following the standard procedure in QED, we now proceed to regularize the
Green's functions by modifying its high-frequency components. Two
traditional methods of regularization are the dimensional and Pauli-Villars
techniques, but neither of these are fully adequate for the problem at hand.
Dimensional regularization is usually applied after an analytic continuation
to the Euclidean path integral, but here we work with the real-time path
integral and nonequilibrium boundary conditions. It turns out that the
Pauli-Villars method does not suppress high-energy field modes strongly
enough for our purposes because it potentially allows high-energy
corrections to modify the long-time particle motion.

Ultimately, there is no reason to prefer any one regularization scheme over
another beyond issues of suitability and particularities (or for preserving
some particular features or symmetries of the high-energy theory). For our
purposes, a sufficiently strongly regulated retarded Green's function is
given by the Gaussian smeared function

\begin{equation}
G_{\Lambda}^{R}(\sigma)=\theta(z^{0}\left( \tau\right) -z^{0}\left(
\tau^{\prime}\right) )\theta\left( \sigma\right) \frac{\Lambda
^{2}e^{-\Lambda^{4}\sigma^{2}/2}}{\sqrt{2\pi^{3}}}
\label{regulated green's function}
\end{equation}
with support inside and on the future lightcone of $z\left( \tau^{\prime
}\right) .$ We have defined 
\begin{align}
\sigma\left( s\right) & =y^{\mu}\left( s\right) y_{\mu}\left( s\right) \\
y^{\mu}\left( s\right) & \equiv z^{\mu}\left( \tau\right) -z^{\mu}\left(
\tau^{\prime}\right)  \nonumber \\
s & \equiv\tau^{\prime}-\tau.  \nonumber
\end{align}
In the limit $\Lambda\rightarrow\infty,$ 
\begin{equation}
\lim_{\Lambda\rightarrow\infty}G_{\Lambda}^{R}\left( \sigma\right)
=\delta\left( \sigma\right) /2\pi,
\end{equation}
giving back the unregulated Green's function. Expanding the function $\sigma$
for small $s$ gives 
\begin{align}
\sigma\left( s\right) & =y^{\mu}\left( s\right) y_{\mu}\left( s\right) \\
& =s^{2}-\ddot{z}^{4}s^{4}/12+{\cal O}\left( s^{5}\right) ,  \nonumber
\end{align}
and 
\begin{equation}
G_{\Lambda}^{R}\left( s\right) =\frac{\Lambda^{2}e^{-\Lambda^{4}s^{4}/2}}{%
\sqrt{2\pi^{3}}},
\end{equation}
assuming a timelike trajectory $z^{\mu}\left( \tau\right) .$

Now that we have defined a cutoff version of QED, we briefly reconsider the
type of initial state assumed in this series of papers. The exact influence
functional is generally found for initially uncorrelated particle(s) and
field, where the initial density matrix is 
\begin{equation}
\hat{\rho}\left( t_{i}\right) =\hat{\rho}_{z}\left( t_{i}\right) \otimes\hat{%
\rho}_{\varphi}\left( t_{i}\right) .  \label{initial state}
\end{equation}
Such a product state, constructed out of the basis $|{\bf z,}%
t\rangle\otimes|\varphi\left( x\right) \rangle,$ is highly excited with
respect to the interacting theory's true ground state. If the theory were
not ultraviolet regulated, producing such a state would require infinite
energy. This is not surprising since it is impossible to take a fully
dressed particle state and strip away all arbitrarily-high-energy
correlations by some finite energy operation. With a cutoff, (\ref{initial
state}) is now a finite energy state and thus physically achievable, though
depending on the cutoff $\Lambda$ it may still require considerable energy
to prepare. The important point, which we show below, is that the late-time
behavior is largely insensitive to the details of the initial state, while
the early-time behavior is fully causal and unique.

\subsection{Single-particle scalar-ALD equation}

The semiclassical limit is given by the equations of motion for the
mean-trajectory. This is just the solution to the Langevin equation (\ref
{Single particle Langevin 1}) without the noise term where, as we indicated
earlier, $\tau$ is the classical proper-time such that $\dot{z}^{2}=N^{2}.$
The regulated semiclassical equations of motion are then

\begin{align}
& m_{0}\ddot{z}_{\mu}\left( \tau\right) -\partial_{\mu}V(z)=C_{1\mu
}^{\left( \eta\right) }  \label{integral equation} \\
& =e^{2}\int_{\tau_{i}}^{\tau}d\tau^{\prime}\left( \ddot{z}_{\mu}\left(
\tau\right) +\dot{z}^{\nu}\left( \tau\right) \dot{z}_{[\mu}\left(
\tau\right) \partial_{\nu]}\right) G_{\Lambda}^{R}(z(\tau),z(\tau^{\prime }))
\nonumber
\end{align}
where the $\partial_{\nu}$ acts only on the $z\left( \tau\right) $ in $%
G_{\Lambda}^{R},$ and not the $z\left( \tau^{\prime}\right) $ term. We add
the subscript to $m_{0}$ because it is the particle's bare mass. This
nonlocal integrodifferential equation is manifestly causal by virtue of the
retarded Green's function. Eq. (\ref{integral equation}) describes the full
non-Markovian semiclassical particle dynamics, which depend on the past
particle history for proper-times in the range $\left[ \tau,\tau_{i}\right]
. $ These second order integrodifferential equations have unique solutions
determined by the ordinary Newtonian (i.e. position and velocity) initial
data at $t_{i}$.

Therefore, the nonlocal equations of motion found from (\ref{integral
equation}) do not have the problems associated with the classical
Abraham-Lorentz-Dirac equation. But, as an integral equation, (\ref{integral
equation}) depends on the entire history of $z\left( \tau\right) $ between $%
\tau_{i}$ and $\tau.$ For this reason, the transformation from (\ref
{integral equation}) to a local differential equation of motion requires
every derivative of $z\left( \tau\right) $ (e.g., $d^{n}z\left( \tau\right)
/d\tau^{n}$). This is the origin of higher-time-derivatives in the local
equations of motion for radiation-reaction (classically as well as
semiclassically). By integrating out the field variables, we have removed an
infinite set of nonlocal (field) degrees of freedom in favor of nonlocal
kernels whose Taylor expansions give higher-derivative terms.

We may now transform the integral equation (\ref{integral equation}) to a
local differential equation by expanding the functions $y^{\mu}\left(
s\right) $ around $s=0.$ In Figure 1, we show a semiclassical trajectory
with respect to the lightcone at $z\left( \tau\right) .$ Taking $\tau$ as
fixed, we change integration variables using $ds=d\tau^{\prime}.$ Next, we
need the expansions 
\begin{align}
-y^{\mu}\left( s\right) & =\dot{z}^{\mu}\left( \tau\right) s+\ddot {z}%
^{\mu}\left( \tau\right) s^{2}/2  \label{rel 1} \\
& +\stackrel{...}{z}^{\mu}\left( \tau\right) s^{3}/6+{\cal O}\left(
s^{4}\right) ,  \nonumber \\
-\dot{y}^{\mu}\left( s\right) & =\dot{z}^{\mu}\left( \tau\right) +\ddot{z}%
^{\mu}\left( \tau\right) s  \nonumber \\
& +\stackrel{...}{z}^{\mu}\left( \tau\right) s^{2}/2+{\cal O}\left(
s^{3}\right) ,  \nonumber \\
\sqrt{\sigma\left( s\right) } & =s-\ddot{z}^{2}s^{3}/24+{\cal O}\left(
s^{4}\right) ,  \label{rel 4}
\end{align}
We set $\dot{z}^{2}=1$ for convenience since we are concerned with the
semiclassical limit. Also, we use $\dot{z}^{\mu}\ddot{z}_{\mu}=0,$ and $%
\ddot{z}^{2}=-\dot{z}^{\mu}\stackrel{...}{z}_{\mu}$ to simplify as
necessary. We denote both $dz/d\tau$ by $\dot{z}$ and $dy/ds$ by $\dot{y}.$
Using 
\begin{align}
\frac{\partial\sigma\left( s\right) }{\partial z^{\mu}} & =2y^{\mu}\left(
s\right) \\
\frac{d\sigma\left( s\right) }{ds} & =2y^{\mu}\dot{y}_{\mu}=2s-\ddot{z}%
^{2}s^{3}/2+{\cal O}\left( s^{4}\right)  \nonumber
\end{align}
allows us to write the gradient operator as 
\begin{align}
\partial_{\mu} & =\frac{\partial\sigma}{\partial z^{\mu}}\frac{d}{d\sigma }%
=2y_{\mu}\left( \frac{d\sigma}{ds}\right) ^{-1}\frac{d}{ds} \\
& =\frac{y_{\mu}}{y^{\nu}\dot{y}_{\nu}}\frac{d}{ds}.  \nonumber
\end{align}
We also define 
\begin{equation}
r=\tau-\tau_{i}
\end{equation}
to be the total elapsed proper-time for the particle since the initial time $%
\tau_{i}$. Recall that $\tau_{i}$ is defined by $z^{0}\left( \tau
_{i}\right) =t_{i}$ where $t_{i}$ is the initial time at which the state $%
\hat{\rho}\left( t_{i}\right) =\hat{\rho}_{z}\left( t_{i}\right) \otimes\hat{%
\rho}_{\varphi}\left( t_{i}\right) $ is defined.

These definitions and relations let us write the RHS of (\ref{integral
equation}) as 
\begin{align}
& e^{2}\int_{0}^{r}ds{\Huge (}u_{\mu}^{\left( 1\right)
}G_{\Lambda}^{R}(s)+u_{\mu}^{\left( 1\right) }\frac{s}{2}\frac{d}{ds}%
G_{\Lambda}^{R}(s)  \label{integral expansion} \\
& +u_{\mu}^{\left( 2\right) }\frac{s^{2}}{6}\frac{d}{ds}G_{%
\Lambda}^{R}(s)+...  \nonumber \\
& +u_{\mu}^{\left( n\right) }\frac{s^{n}}{(n+1)!}\frac{d}{ds}%
G_{\Lambda}^{R}(s)+....{\Huge )}.  \nonumber
\end{align}
where 
\begin{align}
u_{\mu}^{\left( 1\right) } & =\ddot{z}_{\mu}\left( \tau\right)
\label{u(n)'s} \\
u_{\mu}^{\left( 2\right) } & =\left( \dot{z}_{\mu}\ddot{z}^{2}+\stackrel{%
\ldots}{z}_{\mu}\right)  \nonumber \\
& \vdots  \nonumber \\
u_{\mu}^{\left( n\right) } & =u_{\mu}^{\left( n\right) }\left( \dot {z},%
\ddot{z},...,d^{n+1}z/d\tau^{n+1}\right)  \nonumber
\end{align}
The functions $u_{\mu}^{\left( n\right) }$ depend on higher-derivatives of
the particle coordinates up through order $n+1.$ We next define the
time-dependent parameters 
\begin{equation}
h\left( r\right) =\int_{0}^{r}ds\,G_{\Lambda}^{R}\left( s\right)
=a\Lambda\left( 1-\frac{\Gamma\left( 1/4,\Lambda^{4}r^{4}/2\right) }{%
\Gamma(1/4)}\right)
\end{equation}
and 
\begin{align}
g^{\left( n\right) }\left( r\right) & =\int_{0}^{r}ds\,\frac{s^{n}}{(n+1)!}%
\frac{d}{ds}G_{\Lambda}^{R}\left( s\right)  \label{g(n)} \\
& =\frac{32^{-(n-2)/4}}{\pi^{3/2}(n+1)!L^{n-2}}\gamma\left( 1+\frac{n}{4}%
,L^{4}r^{4}/2\right) ,  \nonumber
\end{align}
where $\Gamma\left( x\right) $ is the Gamma function, $\Gamma\left(
x,y\right) $ is the incomplete Gamma function, and $\gamma\left( x,y\right)
=\Gamma\left( x\right) -\Gamma\left( x,y\right) .$ The constant $%
a=3\Gamma\left( 5/4\right) /\left( 2^{1/4}\pi^{3/2}\right) \simeq.41$
depends on the details of the high-energy cutoff, but is of order one for
reasonable cutoffs.

The coefficients are bounded for all $r,$ including the limit $r\rightarrow
\infty .$ In fact, 
\begin{equation}
\lim_{r\rightarrow \infty }g^{\left( n\right) }\left( r\right)
=\lim_{n\rightarrow \infty }g^{\left( n\right) }\left( r\right) =0,
\end{equation}
taking either limit. Therefore, we are free to exchange the sum over $n$ and
the integration over $s$ to find a convergent series expansion of (\ref
{integral equation}). The resulting local equations of motion are 
\begin{align}
& m_{0}\ddot{z}_{\mu }\left( \tau \right) -\partial _{\mu }V(z)=C_{1\mu
}^{\left( \eta \right) }  \label{first cumulant} \\
& =e^{2}h\left( r\right) u_{\mu }^{\left( 1\right) }+e^{2}\sum_{n=1}^{\infty
}g^{\left( n\right) }\left( r\right) u_{\mu }^{\left( n\right) }.  \nonumber
\end{align}
From (\ref{u(n)'s}) we see that the $u^{\left( 1\right) }$ terms give
time-dependent mass renormalization; we accordingly define the renormalized
mass as 
\begin{align}
m\left( r\right) & =m_{0}-e^{2}h\left( r\right) -e^{2}g^{\left( 1\right)
}\left( r\right)   \nonumber \\
& m_{0}+\delta m\left( r\right) .
\end{align}
Similarly, the $u^{\left( 2\right) }$ term is the usual third derivative
radiation reaction force from the ALD equation. Thus, the equations of
motion may be written as 
\begin{equation}
m\left( r\right) \ddot{z}_{\mu }\left( \tau \right) -\partial _{\mu
}V(z)=f_{\mu }^{R.R.}\left( r\right)   \label{radiation reaction force}
\end{equation}
where 
\begin{equation}
f_{\mu }^{R.R.}\left( r\right) =e^{2}g^{\left( 2\right) }\left( r\right)
\left( \dot{z}_{\mu }\ddot{z}^{2}+\stackrel{\ldots }{z}_{\mu }\right)
+e^{2}\sum_{n=3}^{\infty }g^{\left( n\right) }\left( r\right) u_{\mu
}^{\left( n\right) }.  \label{semiclassical2}
\end{equation}

These semiclassical equations of motion are almost of the
Abraham-Lorentz-Dirac form, except for the time-dependent renormalization of
the effective mass, the time-dependent (non-Markovian) radiation reaction
force, and the presence of higher than third-time-derivative terms. The mass
renormalization is to be expected, indeed it occurs in the classical
derivation as well. What an initial-value formulation reveals is that
renormalization effects are time-dependent owing to the nonequilibrium
dynamics of particle-field interactions. The reason why renormalization is
viewed as a time-independent procedure is because one usually only deals
with equilibrium conditions or asymptotic in-out scattering problems.

\section{Renormalization, causality, and runaways}

\subsection{Time-dependent mass renormalization}

The time-dependence of the effective mass $m\left( r\right) $ (and
particularly the radiation reaction coefficients $g^{(n)}\left( r\right) $)
play crucial roles in the semiclassical behavior. The initial particle at
time $t_{i}$ $\left( r=0\right) $ is assumed, by our choice of initial
state, to be fully uncorrelated with the field. This means the particle's
initial mass is just its bare mass $m_{0}.$ Interactions between the
particle and field then re-dress the particle state, one of the consequences
being that the particle acquires a cloud of virtual field excitations
(quanta) that change its effective mass$.$ For a particle coupled to a
scalar field there are two types of mass renormalization effects, one coming
from the $\ddot{z}_{\mu }\varphi $ interaction, and the other coming from
the $\dot{z}^{\mu }\dot{z}_{[\nu }\partial _{\mu ]}\varphi $ interaction. We
commented in Section II.C that the latter interaction is essentially a
scalar field version of the electromagnetic coupling, if one defines the
scalar field analog of the field strength tensor 
\[
F_{\nu \mu }^{scalar}=\dot{z}_{[\nu }\partial _{\mu ]}\varphi . 
\]
We find that this interaction contributes a mass shift at late times of 
\begin{equation}
\delta m_{\dot{z}_{[\nu }\partial _{\mu ]}\varphi }=ae^{2}\Lambda /2.
\end{equation}
In Paper III it is shown that $\delta m_{\dot{z}_{[\nu }\partial _{\mu
]}\varphi }$ is just one half the mass shift that is found when the particle
moves in the electromagnetic field. This is consistent with the
interpretation of the electromagnetic field as equivalent to two scalar
fields (one for each polarization); therefore one expects, and finds, twice
the mass renormalization in the electromagnetic case. (For the same reason
we see below that the radiation reaction force, and stochastic noise, are
each also reduced by half compared to the electromagnetic case.)

But the scalar field, unlike the electromagnetic field, also has a $\ddot{z}%
_{\mu }\varphi $ interaction that gives a mass shift $\delta m_{\ddot{z}%
\varphi }=-2\delta m_{\dot{z}_{[\nu }\partial _{\mu ]}\varphi }.$ The
combined mass shift is then negative. At late-times, the renormalized mass
is 
\begin{align}
m_{\infty }& \equiv \lim_{r\rightarrow \infty }\text{ }m\left( r\right) 
\nonumber \\
& =m_{0}+\delta m_{EM}+\delta m_{scalar}  \nonumber \\
& =m_{0}+ae^{2}\Lambda /2-a\Lambda e^{2}  \nonumber \\
& =m_{0}-ae^{2}\Lambda /2.
\end{align}
For a fixed $m_{0},$ if the cutoff satisfies $\Lambda >2m_{0}/ae^{2},$ the
renormalized mass becomes negative and the equations of motion become
unstable. It is well-known that environments with overly large cutoffs can
qualitative change the system dynamics. In the following we will assume that
the bare mass and cutoff are such that $m_{\infty }>0,$ which in turn
implies that $m_{0}>0$ and gives a bound on the cutoff: 
\[
\Lambda <2m_{0}/ae^{2} 
\]
(which is of the order of the inverse classical electron radius). We
emphasize that for the scalar field case the field interaction{\it \ reduces 
}the effective particle mass, whereas in the electromagnetic (vector) field
case the field interaction {\it increases} the effective particle mass (see
paper III \cite{JH3}). If we followed the FOL \cite
{FOL,FO(runaways),FO(FDR),FO(relativistic)} suggestion of picking a cutoff $%
\Lambda =2m_{0}/ae^{2}$ we would then conclude that $m_{\infty }=0$ as a
consequence of the opposite sign in the mass renormalization effect. This is
a kind of ``critical'' case balanced between stable and unstable particle
motion, and while there may be special instances where such behavior is of
interest, it does not seem to represent the generic behavior of typical
(massive) particles interacting with scalar field degrees of freedom. We
will defer further comparison with EM quantum Langevin equations to Paper
III \cite{JH3} where the mass renormalization has the more familiar sign.

We should further elaborate, however, on the role of causality in the
equations of motion. In the preceding paragraph we see that for the
long-term motion to be stable (i.e. positive effective particle mass) there
is an upper bound on the field cutoff. This is well-known, as detailed in 
\cite{FOL,FO(runaways),FO(FDR),FO(relativistic)}, for example. There, the
QLE is assumed to be of the form 
\begin{equation}
m\ddot{x}+\int_{-\infty}^{t}dt^{\prime}\mu\left( t-t^{\prime}\right) \dot {x}%
\left( t^{\prime}\right) +V^{\prime}\left( x\right) =F\left( t\right) ,
\label{Traditional QLE}
\end{equation}
where the time integration runs from $-\infty$ to $t.$ Causality is related
to the analyticity of $\tilde{\mu}\left( z\right) $ (the Fourier transform
of $\mu\left( t-t^{\prime}\right) $) in the upper half plane of the complex
variable $z.$ The traditional QLE of the type (\ref{Traditional QLE}) with
the lower integration limit set as $t_{i}=-\infty$ effectively makes the
assumption that the particle has been in contact with its environment far
longer than the bath memory time. In our analysis the lower limit in (\ref
{integral equation}) is $\tau_{i},$ not $-\infty$, because it is the
(nonequilibrium) dynamics at early times immediately after the initial
particle/field state is prepared where we have something new to say about
how causality is preserved, and runaways are avoided.

The proper-time dependence of the mass renormalization is shown in Figure 2.
The horizontal axis marks the proper time in units of the cutoff timescale $%
1/\Lambda .$ The vertical scale is arbitrary, depending on the particle
mass. Notice one unusual feature: the mass shift is not monotonic with time,
but instead first overshoots its final asymptotic value. This occurs because
there are two competing mass renormalizing interactions that have slightly
different timescales. The time-dependent mass-shift is a rapid effect, the
final dressed mass is reached within a few $1/\Lambda .$

\subsection{Nonequilibrium radiation reaction}

The constants $g_{n}\left( r\right) $ determine how quickly the particle is
able to rebuild its own self-field, which in turn controls the backreaction
on the particle motion, ensuring that it is causal. We have found, as is
typical of effective field theories, that the equations of motion involve
higher derivative terms (i.e. $d^{n}z/d\tau^{n}$ for $n>3$) beyond the usual
ALD form. In fact, all higher derivative terms are in principle present so
long as they respect the fundamental symmetries (e.g., Lorentz and
reparametrization invariance) of our particle-field model. The timescales
and relative contributions of these higher-derivative forces are determined
by the coefficients $g_{n}\left( r\right) .$ For a fixed cutoff $\Lambda,$
the late-time behavior of the $g_{n}\left( \tau\right) $ scale as

\begin{equation}
\mathrel{\mathop{g_{n}\left( \tau\right) =}\limits_{r\rightarrow\infty}}%
\frac{2^{n/2}\Gamma\left( 1+n/2\right) }{\left( 2\pi\right) ^{3/2}}%
\Lambda^{2-n}
\end{equation}
Therefore, the $g_{2}$ term has the late-time limit 
\begin{equation}
\mathrel{\mathop{g_{2}\left( \tau\right) =}\limits_{r\rightarrow\infty}}%
\frac{1}{4\pi},
\end{equation}
which is independent of $\Lambda.$ In the terminology of effective field
theory, this makes the $n=2$ term renormalizable and implies that it
corresponds to a `marginal' coupling or interaction. In the same terminology
the mass renormalization, scaling as $\Lambda,$ corresponds to a relevant
(renormalizable) coupling. In this vein, the higher-derivative corrections
proportional to $g_{n>2}$, suppressed by powers of the high-energy scale $%
\Lambda,$ correspond to so-called irrelevant couplings.

If we had written the most general action for the particle-field system
consistent with Lorentz and reparametrization invariance, we should have
included from the beginning higher-derivative interactions. The effective
field theory methodology when carried out fully shows that these
high-derivative interactions are in fact irrelevant at low-energies.
Therefore, the ordinary ALD form for radiation reaction (e.g., the $n=2$
term) results not as a consequence of the details of the `true' high-energy
theory, but because any Lorentz and reparametrization invariant effective
theory behaves like a renormalizable particle-field theory that has the ALD\
equation as the low-energy limit. With this we come to the important
conclusion that for a wide class of theories with the appropriate symmetries
the low-energy limit of radiation reaction for (center of mass) particle
motion is given by the ALD result. For theories with additional structure
(such as spin), there will of course be additional radiation reaction terms
important at the energy scale associated with the additional structure.

In Figure 3 we show the proper-time dependence of $g^{\left( 2\right)
}\left( r\right) $, with the same horizontal timescale as in Figure 2. It is
evident that the radiation reaction force approaches the ALD form quickly,
essentially on the cutoff time. While the radiation reaction force is
non-Markovian, the memory time of the field-environment is very short. The
non-Markovian evolution is characterized by the dependence of the
coefficients $m\left( r\right) $ and $g^{\left( 2\right) }\left( r\right) $
on the initial time $\tau_{i},$ but become effectively independent of $%
\tau_{i}$ after $\tau_{\Lambda}\sim1/\Lambda.$ This is an example of the
well-known behavior for quantum Brownian motion models found in \cite{QBM1}.
The effectively transient nature of the short-time behavior helps justify
our use of an initially uncorrelated (factorizable) particle and field state.

Also notice the role played by the particle's elapsed proper-time, $%
r=\tau-\tau_{i}.$ For particles with small average velocities\footnote{%
The small velocity approximation is relative to a choice of reference frame.
In paper I (see Appenix A) we discuss how the choice of initially factorized
state at some time $t_{i}$ picks out a special frame.} the particle's
proper-time is roughly the background Minkowski time coordinate $t.$ In this
case, the time-dependences of the renormalization effects are approximately
the same as one would find in the non-relativistic situation, since $%
r=\left( \tau-\tau_{i}\right) \approx\left( t-t_{i}\right) $. For rapidly
moving particles time-dilation can significantly lengthen the observed
timescales with respect to the background Minkowski time. This indicates
that highly relativistic particles will take longer (with respect to
background time $t)$ to equilibrate with the quantum field/environment than
do more slowly moving particles. Finally, we note that the radiation
reaction force is exactly half that found for the electromagnetic field.

\subsection{Causality and early-time behavior}

We have shown that at low-energies and late-times the ALD radiation reaction
force is generic (neglecting other particle structure, such as spin, which
gives rise to additional low-energy corrections). But the classical ALD
equation is plagued with pathologies like acausal and runaway solutions. The
equations of motion in the form (\ref{integral equation}) are clearly unique
and causal, they require only ordinary initial data, and do not have runaway
solutions. However, it is instructive to see how the equations of motions in
the form (\ref{radiation reaction force},\ref{semiclassical2}), involving
higher derivatives, preserve the causal nature of the solutions with
radiation reaction. To address these questions we now examine the early-time
behavior of the equations of motion.

The presence of higher derivatives in (\ref{semiclassical2}) is an
inevitable consequence of converting from a set of nonlocal integral
equation (\ref{integral equation}) to a set of local differential equations (%
\ref{radiation reaction force},\ref{semiclassical2}). As differential
equations, (\ref{radiation reaction force},\ref{semiclassical2}) may seem to
be unphysical at first-sight because of the apparent need to specify initial
data $z_{\mu }^{\left( n\right) }\left( \tau _{i}\right) =d^{n}z_{\mu
}\left( \tau _{i}\right) /d\tau ^{n}$ for $n\geq 2.$ But the coupling
constants $g_{n}\left( r\right) $ satisfy the crucial property that 
\begin{equation}
g_{n}\left( \tau _{i}\right) =0,
\end{equation}
for all $n.$ We see this graphically for $g_{2}\left( r\right) $ in Figure
3, and analytically for all $g_{n}\left( r\right) $ in (\ref{g(n)}).
Consequently, the particle self-force at $\tau =\tau _{i}$ identically
vanishes (to all orders), and only smoothly rebuilds as the particle's
self-field is reconstituted. Therefore, the initial data at $\tau _{i}$ is
fully (and uniquely) determined by 
\begin{equation}
m\ddot{z}_{\mu }\left( \tau _{i}\right) =-\partial _{\mu }V\left( z\left(
\tau _{i}\right) \right) \text{,}
\end{equation}
which only requires the ordinary Newtonian initial data. The initial values
for the higher derivative terms (e.g. $n\geq 2$) in (\ref{semiclassical2})
are determined iteratively from 
\begin{equation}
m\frac{d^{n}z_{\mu }\left( \tau _{i}\right) }{d\tau _{i}^{n}}=-\frac{d^{n-2}%
}{d\tau _{i}^{n-2}}\partial _{\mu }V\left( z\left( \tau _{i}\right) \right) .
\label{Higher derivative intial data}
\end{equation}
Given the (Newtonian) initial data, the equations of motion are determined
uniquely for all later times, to any order in $n.$ In the classical ALD
equations, one finds runaways even in the case of vanishing external
potential, $V=0.$ In our case, because $g_{n}\left( \tau _{i}\right) =0$ for
all $n,$ if $V=0,$ $z_{\mu }^{\left( n\right) }\left( \tau _{i}\right) =0$
for all $n\geq 2.$ With these initial conditions (and $V=0),$ the equations
of motion (\ref{radiation reaction force},\ref{semiclassical2}) are 
\begin{equation}
\ddot{z}_{\mu }\left( \tau \right) =0
\end{equation}
with unique solutions 
\begin{equation}
z_{\mu }\left( \tau \right) =z_{\mu }\left( \tau _{i}\right) +\left( \tau
-\tau _{i}\right) \dot{z}_{\mu }\left( \tau _{i}\right) .
\end{equation}
Runaways do not arise.

As in the classical derivation of the Abraham-Lorentz-Dirac equation, these
semiclassical equations of motion are equivalent to a second order
differential-integral equation. In the classical case the integral equations
are ``derived'' from the third-order ALD equation. If one chooses boundary
conditions for the integral equation that eliminate runaways, one inevitably
introduces pre-accelerated solutions. In our case, we derive the
higher-derivative equations of motion (\ref{radiation reaction force},\ref
{semiclassical2}) from the explicitly casual non-local integral equation (%
\ref{integral equation}). In doing this we avoid introducing unphysical
solutions at short times as a consequence of the initially vanishing
time-dependent parameters $g_{n}\left( r\right) .$ This turns out to be true
for any good regularization of the field's Green functions.

\section{The stochastic regime and the ALD-Langevin equation}

\subsection{Single particle stochastic limit}

The nonlinear Langevin equation in (\ref{Single particle Langevin 1}) shows
a complex relationship between noise and radiation reaction. A nonlinear
Langevin equation similar to these have been found in stochastic gravity by
Hu and Matacz \cite{HM95EinsteinLangevin} describing the stochastic behavior
of the gravitational field in response to quantum fluctuations of the
stress-energy tensor. We now apply the results in Paper I (see Section IV 
\cite{JH1}) giving the linearized Langevin equation for fluctuations around
the mean trajectory. In the Langevin equation (\ref{Single particle Langevin
1}) we define $z^{\mu}=\bar{z}^{\mu}+\tilde{z}^{\mu},$ where $\bar{z}^{\mu}$
is the semiclassical solution from the previous section, and we work to
first order in the fluctuation variable $\tilde{z}^{\mu}.$ The free kinetic
term is given by 
\begin{align}
& \int d\tau^{\prime}\tilde{z}^{\nu}\left( \tau^{\prime}\right) \left( \frac{%
\delta^{2}S_{z}\left[ \bar{z}\right] }{\delta\bar{z}^{\mu}(\tau )\delta\bar{z%
}^{\nu}(\tau^{\prime})}\right)  \nonumber \\
& =m\left( r\right) \int d\tau^{\prime}\tilde{z}^{\nu}\left( \tau^{\prime
}\right) g_{\mu\nu}\frac{d^{2}}{d\tau^{2}}\delta(\tau-\tau^{\prime }) 
\nonumber \\
& =m\left( r\right) \frac{d^{2}\tilde{z}_{\mu}\left( \tau\right) }{d\tau^{2}}%
.  \label{Langevin Kinetic term}
\end{align}
where we have included the time-dependent mass renormalization effect in the
kinetic term. The external potential term is 
\begin{equation}
\int d\tau^{\prime}\tilde{z}^{\nu}\left( \tau^{\prime}\right) \left( \frac{%
\delta^{2}V(\bar{z})}{\delta\bar{z}^{\mu}(\tau)\delta\bar{z}%
^{\nu}(\tau^{\prime})}\right) =\tilde{z}_{\mu}\left( \tau\right) \frac {%
\partial^{2}V(\bar{z})}{\partial\bar{z}^{\mu2}},
\end{equation}
hence, the second derivative of $V$ acts as a force linearly coupled to $%
\tilde{z}.$ The dissipative term for $\tilde{z}$ involves the (functional)
derivative of the first cumulant (see Eq. (\ref{first cumulant})) with the
mass renormalization $(n=1)$ term removed, as it has already been included
in the kinetic term (\ref{Langevin Kinetic term}) above. We therefore need
the (functional) derivative with respect to $\bar{z}_{\mu}$ of the radiation
reaction term 
\begin{equation}
f_{\mu}^{R.R.}=e^{2}\sum_{n=2}^{\infty}\frac{g^{\left( n\right) }\left(
r\right) }{n!}u_{\mu}^{\left( n\right) }.
\end{equation}
To ${\cal O}\left( \tilde{z}\right) ,$ we have 
\begin{equation}
f_{\mu}^{R.R.}\left( \tilde{z}\left( \tau\right) \right) =\int d\tau^{\prime}%
\tilde{z}_{\nu}\left( \tau^{\prime}\right) \frac{\delta f_{\mu}^{R.R.}\left( 
\tilde{z}\right) }{\delta\bar{z}_{\nu}\left( \tau^{\prime}\right) }.
\end{equation}
We note that $f^{R.R}\left( \bar{z}\right) $ is a function of the
derivatives $\bar{z}_{\mu}^{\left( n\right) }$ with $n>1,$ therefore we have 
\begin{align}
f_{\mu}^{R.R.}\left( \tilde{z}\left( \tau\right) \right) & =\sum
_{n=1}^{\infty}\tilde{z}_{\nu}^{\left( n\right) }\left( \tau\right) \frac{%
\partial f_{\mu}^{R.R.}\left( \tilde{z}\right) }{\partial\bar{z}_{\nu
}^{\left( n\right) }} \\
& =e^{2}\sum_{n=1}^{m+1}\tilde{z}_{\nu}^{\left( n\right) }\left( \tau\right)
\sum_{m=2}^{\infty}\frac{g^{\left( m\right) }\left( r\right) }{m!}\frac{%
\partial u_{\mu}^{\left( m\right) }}{\partial\bar{z}_{\nu }^{\left( n\right)
}},  \nonumber
\end{align}
where $\partial f^{R.R.}/\partial\bar{z}^{\left( n\right) }$ are partial
derivatives with respect to the $n^{th}$ derivative of $\bar{z}\left(
\tau\right) .$ The sum over $n$ terminates at $m+1$ since $u_{\mu}^{\left(
m\right) }$ only contains derivatives up through order $z^{\left( m+1\right)
}.$ Hence, the Langevin equations for the fluctuations $\tilde{z}$ (around $%
\bar{z})$ are 
\begin{align}
& m_{R}\left( t\right) \frac{d^{2}\tilde{z}_{\mu}\left( \tau\right) }{%
d\tau^{2}}-\tilde{z}_{\mu}\left( \tau\right) \frac{\delta^{2}V(z)}{\delta%
\bar{z}^{\mu2}}  \label{Single particle langevin- all orders} \\
& =\eta_{\mu}\left( \tau\right) +e^{2}\sum_{n=1}^{m+1}\sum_{m=2}^{\infty
}\left( \frac{g^{\left( m\right) }\left( r\right) }{m!}\frac{\partial
u_{\mu}^{\left( m\right) }}{\partial\bar{z}_{\nu}^{\left( n\right) }}\right) 
\tilde{z}_{\nu}^{\left( n\right) }\left( \tau\right) .  \nonumber
\end{align}
These are solved with $\bar{z}$ found from (\ref{radiation reaction force}).
The noise $\eta_{\mu}\left( \tau\right) $ is given by 
\begin{equation}
\eta_{\mu}\left( \tau\right) =\vec{w}_{\mu}\chi\left( \bar{z}\left(
\tau\right) \right) =e\left( \stackrel{..}{\bar{z}}_{\mu}\left( \tau\right) +%
\stackrel{.}{\bar{z}}^{\nu}\stackrel{.}{\bar{z}}_{[\nu}\partial_{\mu]}%
\right) \chi\left( \bar{z}\left( \tau\right) \right) ,
\label{single particle noise}
\end{equation}
where the lowest order noise is evaluated using the semiclassical solution $%
\bar{z}.$

The correlator $\left\langle \eta_{\mu}\left( \tau\right) \eta_{\nu}\left(
\tau^{\prime}\right) \right\rangle $ (see Eq. \ref{noise correlator,single
particle}) is then found using the field correlator along the
mean-trajectory: 
\begin{equation}
\left\langle \left\{ \chi(\bar{z}(\tau)),\chi\left( \bar{z}\left(
\tau^{\prime}\right) \right) \right\} \right\rangle =\hbar G^{H}\left( \bar{z%
}\left( \tau\right) ,\bar{z}\left( \tau^{\prime}\right) \right) .
\end{equation}
Equation (\ref{single particle noise}) shows that the noise experienced by
the particle depends not just on the stochastic properties of the quantum
field, but on the mean-solution $\bar{z}\left( \tau\right) .$ For instance,
the $\stackrel{..}{\bar{z}}_{\mu}\chi(z)$ term in (\ref{single particle
noise}) gives noise that is proportional to the average particle
acceleration. The second term in (\ref{single particle noise}) depends on
the antisymmetrized combination $\stackrel{.}{\bar{z}}_{[\mu}\partial_{\nu]}%
\chi(\bar{z}).$ It follows immediately from $\left\langle \chi\right\rangle
=0$ that $\left\langle \eta\left( \tau\right) \right\rangle =0.$

From (\ref{noise correlator,single particle}), we see that the noise
correlator is 
\begin{align}
& \left\langle \eta_{\mu}\left( \tau\right) \eta_{\nu}\left( \tau^{\prime
}\right) \right\rangle  \nonumber \\
& =\vec{w}_{\mu}(\tau)\vec{w}_{\nu}\left( \tau^{\prime}\right) \left\langle
\chi(\bar{z}\left( \tau\right) )\chi\left( \bar{z}\left( \tau^{\prime
}\right) \right) \right\rangle  \nonumber \\
& =\left( e^{2}\hbar/c^{4}\right) \left( \stackrel{..}{\bar{z}}_{\mu }+%
\stackrel{.}{\bar{z}}^{\lambda}\stackrel{.}{\bar{z}}_{[\mu}\partial_{\lambda
]}^{z}\right) \left( \stackrel{..}{\bar{z}}_{\nu}^{\prime}+\stackrel{.}{\bar{%
z}}^{\rho\prime}\stackrel{.}{\bar{z}}_{[\nu}^{\prime}\partial_{\rho
]}^{z^{\prime}}\right)  \nonumber \\
& \times G^{H}(\bar{z},\bar{z}^{\prime})  \nonumber \\
& =\left( \stackrel{..}{\bar{z}}_{\mu}+\stackrel{.}{\bar{z}}^{\lambda}\frac{%
\stackrel{.}{\bar{z}}_{[\mu}\bar{y}_{\lambda]}}{\bar{y}^{2}}\frac {d}{d\bar{%
\sigma}}\right) \left( \stackrel{..}{\bar{z}}_{\nu}^{\prime }+\stackrel{.}{%
\bar{z}}^{\rho\prime}\frac{\stackrel{.}{\bar{z}}_{[\nu}^{\prime }\bar{y}%
_{\lambda]}}{\bar{y}^{2}}\frac{d}{d\bar{\sigma}}\right)
\label{noise correlator,single particle} \\
& \times\left( 4\pi\lambda_{c}r_{c}\right) G^{H}(\bar{\sigma}).  \nonumber
\end{align}
We have defined $\bar{y}_{\mu}\equiv\bar{z}_{\mu}\left( \tau\right) =\bar {z}%
_{\mu}\left( \tau^{\prime}\right) $ and $\bar{\sigma}=\bar{y}^{2}.$ The
noise (when the field is initially in its vacuum state) is highly nonlocal,
reflecting the highly correlated nature of the quantum vacuum. Only in the
high temperature limit does the noise become approximately local. Notice
that the noise correlator inherits an implicit dependence on the initial
conditions at $\tau_{i}$ through the time-dependent equations of motion for $%
\bar{z}.$ But when $t\Lambda\gg1,$ the equations of motion become
effectively independent of the initial time. The noise given by (\ref{noise
correlator,single particle}) is not stationary except for special cases of
the solution $\bar{z}(\tau).$

Since the $n>2$ terms represent high-energy corrections suppressed by powers
of $\Lambda,$ the low-energy Langevin equations are well approximated by
keeping only the $n=2$ term. The linear dissipation ($n=2$) term becomes,
using $u_{\mu}^{\left( 2\right) }=(\stackrel{.}{\bar{z}}_{\mu}\stackrel{..}{%
\bar{z}}^{2}+\stackrel{...}{\bar{z}}_{\mu}),$ 
\begin{align}
& e^{2}\sum_{n=1}^{3}\tilde{z}_{\nu}^{\left( n\right) }\left( \tau\right)
\left( \frac{g^{\left( 2\right) }\left( r\right) }{2!}\frac{\partial
u_{\mu}^{\left( 2\right) }}{\partial\bar{z}_{\nu}^{\left( n\right) }}\right)
\\
& =\frac{e^{2}}{2}g^{\left( 2\right) }\left( r\right) (\stackrel{.}{\tilde{z}%
}^{\nu}\stackrel{..}{\bar{z}}^{2}\left( g_{\mu\nu}-\stackrel{.}{\bar{z}}%
_{\mu}\stackrel{...}{\bar{z}}_{\nu}\right)  \nonumber \\
& +\stackrel{...}{\tilde{z}}^{\nu}\left( g_{\mu\nu}-\stackrel{.}{\bar{z}}%
_{\mu }\stackrel{.}{\bar{z}}_{\nu}\right) )  \nonumber \\
& =\frac{e^{2}}{2}g^{\left( 2\right) }\left( r\right) \left(
a^{2}h_{\mu\nu}^{\left( 1\right) }\stackrel{.}{\tilde{z}}^{\nu}+h_{\mu\nu
}^{\left( 2\right) }\stackrel{...}{\tilde{z}}^{\nu}\right) ,  \nonumber
\end{align}
where 
\begin{align}
h_{\mu\nu}^{\left( 1\right) } & \equiv\left( g_{\mu\nu}-\stackrel{.}{\bar{z}}%
_{\mu}\stackrel{...}{\bar{z}}_{\nu}\right) \\
h_{\mu\nu}^{\left( 2\right) } & \equiv\left( g_{\mu\nu}-\stackrel{.}{\bar{z}}%
_{\mu}\stackrel{.}{\bar{z}}_{\nu}\right)  \nonumber
\end{align}
and $a^{2}=\stackrel{..}{\bar{z}}^{2}.$ This lowest order expression for
dissipation in the equations of motion for $\tilde{z}_{\mu}$ involves time
derivatives of $\tilde{z}$ through third order. However, just as was the
case for the semiclassical equations of motion, the vanishing of $g^{\left(
2\right) }\left( r\right) $ for $r=0$ implies that the initial data for the
equations of motion are just the ordinary kind involving no higher than
first order derivatives. The lowest-order late-time ($\Lambda r>>1)$ single
particle Langevin equations are thus 
\begin{align}
\eta\left( \tau\right) & =m\stackrel{..}{\tilde{z}}_{\mu}\left( \tau\right) +%
\tilde{z}_{\nu}\frac{\partial V\left( \bar{z}\right) }{\partial\bar{z}%
^{\mu}\partial\bar{z}_{\nu}} \\
& -\frac{e^{2}}{8\pi}\left( a^{2}h_{\mu\nu}^{\left( 1\right) }\stackrel{.}{%
\tilde{z}}^{\nu}+h_{\mu\nu}^{\left( 2\right) }\stackrel{...}{\tilde{z}}%
^{\nu}\right) .  \nonumber
\end{align}

\subsection{Multiparticle stochastic limit and stochastic Ward identities}

It is a straightforward generalization to construct multiparticle Langevin
equations. The two additional features are particle-particle interactions,
and particle-particle correlations. The semiclassical limit is modified by
the addition of the terms 
\begin{equation}
e^{2}\sum_{m\neq n}\int_{\tau_{i}}^{\tau_{f}}d\tau_{m}\vec{w}_{n}^{\mu}\left[
z_{n}\left( \tau_{n}\right) )G_{\Lambda}^{R}(z_{n}(\tau_{n}),z_{m}(\tau
_{m}))\right] .  \label{particle-particle interaction}
\end{equation}
The use of the regulated $G_{\Lambda}$ is essential for consistency between
the radiation that is emitted during the regime of time-dependent
renormalization and radiation-reaction; it ensures agreement between the
work done by radiation-reaction and the radiant energy. A field cutoff $%
\Lambda$ implies that the radiation wave front emitted at $\tau_{i}$ is
smoothed on a time scale $\Lambda^{-1}$.

We note one significant difference between the single particle and
multiparticle theories. For a single particle, the dissipation is local
(when the field is massless), and the semiclassical limit (obtaining when $%
r\Lambda\gg1)$ is essentially Markovian. The multiparticle theory is
non-Markovian even in the semiclassical limit because of multiparticle
interactions. Particle A may indirectly depend on its own past state of
motion through the emission of radiation which interacts with another
particle B, which in turn, emits radiation that re-influences particle A at
some point in the future. The nonlocal field degrees of freedom stores
information (in the form of radiation) about the particle's past. Only for a
single timelike particle in flat space without boundaries is this
information permanently lost insofar as the particle's future motion is
concerned. These well-known facts make the multiparticle behavior extremely
complicated.

We may evaluate the integral in (\ref{particle-particle interaction}) using
the identity 
\begin{equation}
\delta\left( \left( z_{n}-z_{m}\right) ^{2}\right) =\delta(\tau-\tau
^{\ast})/2\left( z_{n}-z_{m}\right) ^{\nu}\dot{z}_{m\nu,}
\end{equation}
where $\tau^{\ast}$ is the time when particle $m$ crosses particle $%
n^{\prime }s$ past lightcone. (The regulated Green's function would spread
this out over approximately a time $\Lambda^{-1}).$ Defining $z_{nm}\equiv
z_{n}-z_{m}$, we use 
\begin{align}
& \int d\tau_{m}\partial_{m}^{\lambda}G^{R}\left( z_{n},z_{m}\right) \\
& =-\int d\tau_{m}\frac{dG^{R}}{ds}\frac{\left( z_{nm}\right) ^{\lambda}}{%
\left( z_{nm}\right) ^{\alpha}\dot{z}_{m\alpha}}  \nonumber \\
& =-\left[ \frac{G^{R}\left( z_{nm}\right) ^{\lambda}}{\left( z_{nm}\right)
^{\alpha}\dot{z}_{m\alpha}}\right] +\int dsG^{R}\frac{d}{ds}\left[ \frac{%
\left( z_{nm}\right) ^{\lambda}}{\left( z_{nm}\right) ^{\alpha}\dot{z}%
_{m\alpha}}\right]  \nonumber \\
& =\frac{1}{2\left( z_{nm}\right) ^{\nu}\dot{z}_{m\nu}}\frac{d}{ds}\left[ 
\frac{\left( z_{nm}\right) ^{\lambda}}{\left( z_{nm}\right) ^{\alpha }\dot{z}%
_{m\alpha}}\right] .  \nonumber
\end{align}
The particle-particle interaction terms are then given by 
\begin{align}
& e^{2}\sum_{m\neq n}\left( \ddot{z}_{n}^{\mu}+\dot{z}_{\beta}\dot{z}^{[\mu
}\partial^{\beta]}\right) G^{R}\left( z_{n},z_{m}\right)
\label{Lienard-Wiechart: scalar} \\
& =e^{2}\sum_{m\neq n}\left( \frac{\ddot{z}_{n}^{\mu}}{\left[ \left(
z_{nm}\right) ^{\nu}\dot{z}_{m\nu}\right] }\right.  \nonumber \\
& +\left. \frac{\dot{z}_{n}^{\mu}\dot{z}_{n\lambda}-\delta_{\lambda}^{\mu}}{%
2\left( z_{nm}\right) ^{\nu}\dot{z}_{m\nu}}\frac{d}{d\tau_{m}}\left[ \frac{%
\left( z_{nm}\right) ^{\lambda}}{\left( z_{nm}\right) ^{\alpha }\dot{z}%
_{m\alpha}}\right] \right) _{\tau_{m}=\tau^{\ast}},  \nonumber
\end{align}
where, as before, the $\vec{w}_{n}^{\mu}\left( z_{n}\right) $ acts only on
the $z_{n},$ and not the $z_{m}\left( \tau^{\ast}\right) .$ Eq. (\ref
{Lienard-Wiechart: scalar}) are just the scalar analog of the
Li\'{e}nard-Wiechert forces. They include both the near-field and far-field
effects.

The long range particle-particle terms in the Langevin equations are found
using (\ref{nonlinear multiparticle langevin equation}), so we have the
additional Langevin term for the $n$th particle, found from 
\begin{align}
& e^{2}\sum_{m\neq n}\int d\tau_{n}\tilde{z}_{n}^{\nu}\partial_{\nu}^{n}%
\left( \frac{\stackrel{..}{\bar{z}}_{n}^{\mu}}{\left[ \left( \bar {z}%
_{nm}\right) ^{\alpha}\stackrel{.}{\bar{z}}_{m\alpha}\right] }\right.
\label{multiparticle scalar interactions} \\
& -\left. \frac{h_{n,\mu\lambda}^{\left( 2\right) }}{2\left( \bar{z}%
_{nm}\right) ^{\alpha}\stackrel{.}{\bar{z}}_{m\alpha}}\frac{d}{d\tau_{m}}%
\left[ \frac{\left( \bar{z}_{nm}\right) ^{\lambda}}{\left( \bar{z}%
_{nm}\right) ^{\beta}\stackrel{.}{\bar{z}}_{m\beta}}\right] \right)
_{\tau_{m}=\tau^{\ast}}.  \nonumber
\end{align}
Equation (\ref{multiparticle scalar interactions}) contains no third (or
higher) derivative terms. The multiparticle noise correlator is 
\begin{equation}
\left\langle \left\{ \eta_{n}^{\mu}\left( \tau_{n}\right) \eta_{m}^{\nu
}\left( \tau_{m}\right) \right\} \right\rangle =\vec{w}_{n}^{\mu}\left(
\tau_{n}\right) \vec{w}_{m}^{\nu}\left( \tau_{m}\right) \left\langle \left\{
\chi\left( \bar{z}_{n}\right) \chi\left( \bar{z}_{m}\right) \right\}
\right\rangle .
\end{equation}
In general, particle-particle correlations between spacelike separated
points will not vanish as a consequence of the nonlocal correlations
implicitly encoded by $G^{H}.$ For instance, when there are two oppositely
charged particles which never enter each other's causal future, there will
be no particle-particle interactions mediated by $G^{R};$ however, the
particles will still be correlated through the field vacuum via $G^{H}$.
This shows that it will not be possible to find a generalized multiparticle
fluctuation-dissipation relation under all circumstances \cite
{RavalHu:StochasticAcceleratedDetectors}.

We now briefly consider general properties of the stochastic equations of
motion. Notice that the noise satisfies the identity 
\begin{equation}
\dot{z}_{n}^{\mu}\left( \tau\right) \eta_{n\mu}\left( \tau\right) =\dot {z}%
_{n}^{\mu}\left( \tau\right) \vec{w}_{_{n}\mu}(z_{n})\chi(z_{n})=0,
\label{z dot n vanishes}
\end{equation}
which follows as a consequence of $\dot{z}_{n}^{\mu}\vec{w}_{n\mu}=0$, for
each particle number separately. This is an essential property since the
particle fluctuations are real, not virtual. We may use (\ref{z dot n
vanishes}) to prove what might be called (by analogy with QED) stochastic
Ward-Takahashi identities \cite{Ward}. The n-point correlation functions for
the particle-noise are 
\begin{align}
& \left\langle \left\{ \eta_{\mu_{1}}(\tau_{1})...\eta_{\mu_{i}}\left(
\tau_{i}\right) ...\eta_{\mu_{n}}\left( \tau_{n}\right) \right\}
\right\rangle  \label{n-point correlation functions} \\
& =\vec{w}_{\mu_{1}}\left( z_{1}\right) ...\vec{w}_{\mu_{i}}\left(
z_{i}\right) ...\vec{w}_{\mu_{n}}\left( z_{n}\right)  \nonumber \\
& \times\left\langle \left\{ \chi\left( z_{1}\right) ...\chi\left(
z_{i}\right) ...\chi\left( z_{n}\right) \right\} \right\rangle ,  \nonumber
\end{align}
with each $\vec{w}_{\mu_{i}}\left( z_{i}\right) $ acting on the
corresponding $\chi(z_{i}).$ These correlation functions may both involve
different times along the worldline of one particle (i.e. self-particle
noise), and correlations between different particles. From (\ref{n-point
correlation functions}) and (\ref{z dot n vanishes}), follow ``Ward''
identities:

\begin{equation}
\dot{z}_{i}^{\mu_{i}}\,\left\langle
\eta_{\mu_{1}}(\tau_{1})...\eta_{\mu_{i}}\left( \tau_{i}\right)
...\eta_{\mu_{n}}\left( \tau_{n}\right) \right\rangle =0,\;\;\;\text{for all 
}i\text{ and }n.  \label{Ward identities}
\end{equation}
The contraction of on-shell momenta ($\dot{z}^{\mu}$) with the stochastic
correlation functions always vanishes. These identities are fundamental to
the consistency of the relativistic Langevin equations.

\section{Example: Free particles in the scalar field vacuum}

As a concrete example, we find the Langevin equations for $V(z)=0$ and the
scalar field initially in the vacuum state. Using the stochastic equations
of motion we can address the question of whether a free particle will
experience Brownian motion induced by the vacuum fluctuations of the scalar
field. When $V(z)=0$, we immediately see that for the semiclassical
equations, $\stackrel{..}{\bar{z}}^{\mu}\left( \tau\right) =$ constant$;$
the radiation-reaction force identically vanishes (for all orders of $n$).
The fact that $g_{n}\left( 0\right) =0$ uniquely fixes the constant as zero
due to the initial data boundary conditions for the higher derivative terms (%
\ref{Higher derivative intial data}); and hence, there are no runaway
solutions characterized by a non-zero constant. We may write $\stackrel{.}{%
\bar{z}}^{\mu}\left( \tau\right) =v^{\mu}$ and $\bar{z}^{\mu}\left(
\tau\right) =sv^{\mu},$ where $v^{\mu}$ is a constant spacetime velocity
vector which satisfies the (mass-shell) constraint $v^{2}=1.$ A particle
moving in accordance with the semiclassical solution neither radiates nor
experiences radiation-reaction\footnote{%
In constast, the mean-solution for a free particle in a quantum scalar field
found in \cite{Gour-Sriramkumar(98)} does not possess the Galilean
invariance of our results, and its motion is furthermore damped
proportionally to its velocity.}. Using $\bar{\sigma}^{2}=v^{2}s^{2}=s^{2}=%
\left( \tau-\tau^{\prime}\right) ^{2},$ we immediately find from (\ref
{Single particle Langevin 1}) and (\ref{Single particle langevin- all orders}%
) (using $a^{2}=\stackrel{..}{\bar{z}}^{2}=0$) the (linear-order) Langevin
equation 
\begin{equation}
m\frac{d^{2}\tilde{z}_{\mu}\left( \tau\right) }{d\tau^{2}}+\frac{e^{2}}{2}%
g^{\left( 2\right) }\left( r\right) \left( g_{\mu\nu}-\stackrel{.}{\bar{z}}%
_{\mu}\stackrel{.}{\bar{z}}_{\nu}\right) \stackrel{...}{\tilde{z}}%
^{\nu}=\eta_{\mu}\left( \tau\right) .  \label{Free particle Langevin}
\end{equation}
The noise correlator is found from (\ref{noise correlator,single particle})
to be 
\begin{align}
& \left\langle \eta_{\eta}\left( \tau\right) \eta_{\nu}\left( \tau^{\prime
}\right) \right\rangle  \label{correlator of single particle noise} \\
& =e^{2}\hbar\left( \stackrel{..}{\bar{z}}_{\mu}-\frac{d}{d\tau}\left( 
\stackrel{.}{\bar{z}}^{\lambda}\frac{\stackrel{.}{\bar{z}}_{[\mu}\bar {y}%
_{\lambda]}}{\bar{y}^{\alpha}\stackrel{.}{\bar{y}}_{\alpha}}\left( \frac{d%
\bar{\sigma}^{2}}{d\tau}\right) ^{-1}\right) \right)  \nonumber \\
& \times\left( 4\pi^{2}\bar{\sigma}^{2}\right) ^{-1}\left( \stackrel{..}{%
\bar{z}}_{\nu}^{\prime}+\frac{d}{d\tau^{\prime}}\left( \stackrel{.}{\bar{z}}%
^{\rho\prime}\frac{\stackrel{.}{\bar{z}}_{[\nu}^{\prime}\bar {y}_{\lambda]}}{%
\bar{y}^{\beta}\stackrel{.}{\bar{y}}_{\beta}}\left( \frac{d\bar{\sigma}^{2}}{%
d\tau^{\prime}}\right) ^{-1}\right) \right)  \nonumber \\
& =e^{2}\hbar\left( \frac{d}{d\tau}\left( v^{\lambda}\frac{v_{[\mu
}v_{\lambda]}s}{2v^{\alpha}v_{\alpha}s^{2}}\right) \right) \left( 4\pi
^{2}s^{2}\right) ^{-1}  \nonumber \\
& \times\left( \frac{d}{d\tau^{\prime}}\left( v^{\rho}\frac{v_{[\nu
}v_{\lambda]s}}{-2v^{\beta}v_{\beta}s^{2}}\right) \right) ,
\label{second term in correator}
\end{align}
where we have substituted the vacuum Hadamard function for the field-noise
correlator $\left\langle \chi\chi^{\prime}\right\rangle .$ The antisymmetric
combination $v_{[\mu}v_{\nu]}$ vanishes, and hence, 
\begin{equation}
\left\langle \eta_{\eta}\left( \tau\right) \eta_{\nu}\left( \tau^{\prime
}\right) \right\rangle =0,\;\;\text{ for }\stackrel{..}{\bar{z}}^{\mu }=0.
\label{inertial particle noise}
\end{equation}
The origin of this at first surprising result is not hard to find. First,
the ``$\stackrel{..}{\bar{z}}\chi$'' dependant terms vanish since the
average acceleration is zero, but why should the other terms in the
correlator vanish? From (\ref{second term in correator}), we see that the
second term in the particle-noise correlator involves $\stackrel{.}{\bar{z}}%
^{\lambda}\stackrel{.}{\bar{z}}_{[\mu}\partial_{\lambda]}^{z}G^{H}(\bar{z},%
\bar{z}^{\prime}).$ But the gradient of the vacuum Hadamard kernel (which is
a function of $\sigma$) satisfies $\partial_{\mu}G^{H}(\sigma)\propto
y_{\mu},$ and is therefore always in the direction of the spacetime vector
connecting $\bar {z}\left( \tau\right) $ and $\bar{z}\left(
\tau^{\prime}\right) .$ However, for the inertial particle, $%
y_{\mu}\varpropto v_{\mu}$, and thus the antisymmetrized term, $\stackrel{.}{%
\bar{z}}_{[\mu}\partial_{\nu]}G^{H}(\bar{\sigma})\propto
v_{[\mu}v_{\nu]}G^{H},$ always vanishes. It therefore follows that a scalar
field with the coupling that we have assumed does not induce stochastic
fluctuations in a free particles trajectory. This result is actually
somewhat a special case for scalar fields, because in Paper III \cite{JH3}
we show that the electromagnetic field does induce stochastic fluctuations
in a free particle. In the case of the electromagnetic field there is more
freedom in $F_{\mu\nu}^{EM}=\partial_{\lbrack\mu}A_{\nu]}$ (e.g., $A_{\mu}$
is a vector field; $\partial_{\mu}\varphi$ is the gradient of a scalar
field) and so one does not find the cancellation that occurs above. For the
case of a scalar field, there are noise-fluctuations when the field is not
in the vacuum state (e.g. a thermal quantum field), and/or when the particle
is subject to external forces that make its average acceleration non-zero
(e.g. an accelerated particle).

We therefore find that the free particle fluctuations (in this special case)
do not obey a Langevin equation. The equations of motion for $\tilde{z}%
_{\mu} $ are 
\begin{equation}
m\frac{d^{2}\tilde{z}_{\mu}\left( \tau\right) }{d\tau^{2}}+\frac{e^{2}}{2}%
g^{\left( 2\right) }\left( r\right) \left( g_{\mu\nu}-\stackrel{.}{\bar{z}}%
_{\mu}\stackrel{.}{\bar{z}}_{\nu}\right) \stackrel{...}{\tilde{z}}^{\nu}=0
\end{equation}
Eq. (\ref{Free particle Langevin}) has the unique solution 
\begin{equation}
\stackrel{..}{\tilde{z}}_{\mu}\left( \tau\right) =0
\end{equation}
after recalling that $g^{\left( 2\right) }\left( 0\right) =0\Rightarrow 
\stackrel{...}{\tilde{z}}^{\mu}\left( 0\right) =0.$ That is, the same
initial time behavior that gives unique (runaway-free) solutions for $\bar{z}
$ (the semiclassical solution) apply to $\tilde{z}$ (the stochastic
fluctuations).

Let us comment on the difference of our result from that of \cite
{Gour-Sriramkumar(98)}. First, the dissipation term in our equations of
motion is relativistically invariant for motion in the vacuum, and vanishes
for inertial motion. The equations of motion in \cite{Gour-Sriramkumar(98)}
are not; the authors concluded that a particle moving through the scalar
vacuum will experience a dissipation force proportional to its velocity in
direct contradiction with experience. This result comes from an incorrect
treatment of the retarded Green's function in 1+1 spacetime dimensions-
which is quite different in behavior than the Green's function in 3+1
dimensions. We avoid this problem by working in the more physical three
spatial dimensions. We furthermore find that a free particle in the scalar
vacuum does not experience noise (at least at the one-loop order of our
derivation- see Paper I). This result is somewhat special owing to the
particular symmetries of the scalar theory. In Paper III, we find that
charged particles experience noise in the electromagnetic field vacuum. We
have not addressed in depth the degree of decoherence of the particle
motion, and how well the Langevin equation approximates the quantum
expectation values for the particle motion is still an open question that
will depend on the particular parameters (e.g., mass, charge, state of the
field) of the model.

Our result also differs from that of Ford and O'Connell, who propose that
the field-cutoff take a special value with the consequence that a free
electron experiences fluctuations without dissipation \cite{FO(FDR)}. We
have already argued why this assumption is unnecessary as a consequence of
recognizing QED as an effective theory. We have also emphasized the
distinction between radiation reaction and dissipation. Hence, we find that
a free (or for that matter a uniformly accelerated) particle has vanishing
average radiation reaction, but when there exists a fluctuating force
inducing deviations from the free (or uniformly accelerated) trajectory,
there is then a balancing dissipation backreaction force. What is a little
unusual about the scalar-field-free-particle case is that the ``overly
simple'' monopole coupling between the scalar field and particle does not
lead to quantum fluctuation induced noise when the expectation value of the
particles acceleration vanishes (i.e. for a free particle). In paper III 
\cite{JH3}, treating the electromagnetic (vector) field, there {\it is}
quantum field induced noise and dissipation in the particle trajectory, for
both free and accelerated motion.

\section{Discussion}

We now summarize the main results of this paper, what follows from here, and
point out areas for potential applications of theoretical and practical
values.

Perhaps the most significant difference between this and earlier work in
terms of approach is the use of a particle-centric and initial value
(causal) formulation of relativistic quantum field theory, in terms of
world-line quantization and influence functional formalisms, with focus on
the coarse-grained and stochastic effective actions and their derived
stochastic equations of motion. This is a general approach whose range of
applicability extends from the full quantum to the classical regime, and
should not be viewed as an approximation scheme valid only for the
semiclassical. There exists a great variety of physical problems where a
particle-centric formulation is more adept than a field-centric formulation.
The initial value formulation with full backreaction ensures the
self-consistency and causal behavior of the equations of motion for the
semiclassical limit of particles in QED whose solutions are pathology free--
the ALD-Langevin equations.

For further development, in Paper III \cite{JH3} we shall extend these
results to spinless particles moving in the quantum electromagnetic field,
where we need to deal with the issues of gauge invariance. We find that the
electromagnetic field is a richer source of noise than the simpler scalar
field treated here. The ALD-Langevin equations for charged particle motion
in the quantum electromagnetic field derived in Paper III is of particular
importance for quantum beam dynamics and heavy-ion physics. Applying the
ALD-Langevin equations to a charged particle moving in a constant background
electric field (with the quantum field in the vacuum state), we show that
its motion experiences stochastic fluctuations that are identical to those
experienced by a free particle (i.e., where there is no background electric
or magnetic field) moving in a thermal quantum field. This is an example of
the Unruh effect \cite{Unruh76(Original-discovery-Unruh-effect)}. The same
effect exists for the scalar field example treated in this work, but we
leave the essentially identical derivation for the more physically relevant
case of QED presented in \cite{JH3}.

In a second series of papers \cite{JH4-5}, we use the same conceptual
framework and methodology but go beyond the semiclassical and stochastic
regimes to incorporate the full range of quantum phenomena, addressing
questions such as the role of dissipation and correlation in charged
particle pair-creation, and other quantum relativistic processes.

In conclusion, our approach is appropriate for any situation where particle
motion (as opposed to field properties, say) is the center of attention. Our
methods can be applied to relativistic charged particle motion not only in
charged particle beams as in accelerators, and free-electron-lasers as in
ion-optics, but also in strong fields such as particles moving in matter
(crystals) or in plasma media as in astrophysical contexts \cite{PisinChen}.

Self-consistent treatment of backreaction from quantum field activities on a
charged particle is an essential yet often neglected factor in many problems
dealing with charged particle motion. First principles microscopic approach
by means of covariant quantum field theory leading to microscopic stochastic
equation is an effective remedy to this deficiency.

Theoretically, by interpreting a particle's spacetime properties (e.g. its
location in time and space) as effectively microscopic `clocks and position
markers' (e.g. events) we see that a quantum-covariant description of
particle `worldline' motion is not just particle dynamics, it is the quantum
dynamics of spacetime events. For example, the ``time'' of the particle is
treated as quantum variable on equal footing with position, and hence
question like localization apply not just to localization in space, but
localization in time. It would be natural to apply this approach to subjects
like `time of flight', that have become popular, and questions regarding
decoherence of time (variables), together with space variables (e.g. the
emergence of (local) time for particles).

\section*{Acknowledgments}

This work is supported by NSF grant PHY98-00967 and DOE grant
DEFG0296ER40949.

%

%
\newpage

\begin{figure}[tbh]
\begin{center}
\epsfig{file=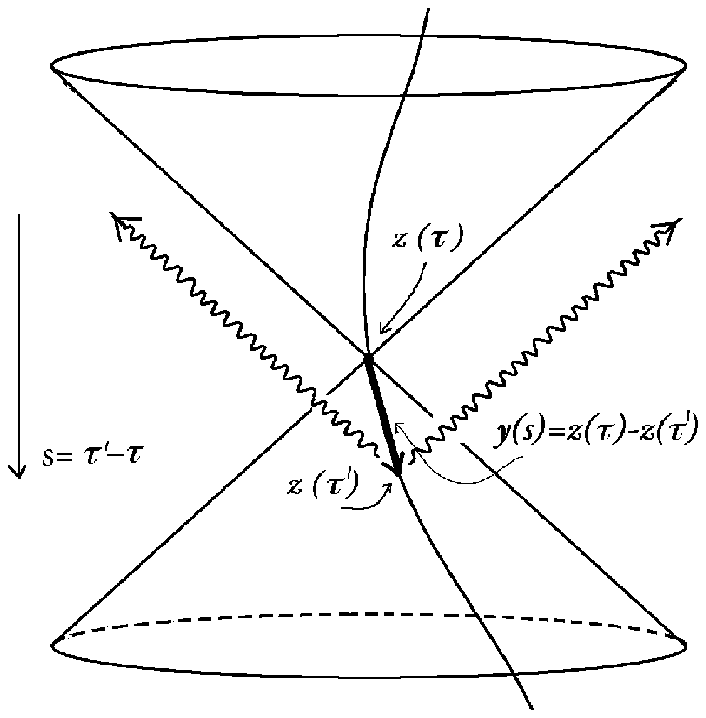, width=3 in}
\end{center}
\caption{The trajectory and the lightcone of a particle at $z\left(
\tau\right)  .$ The radiation reaction force on a particle at $z\left(
\tau\right)  $ depends on the particle trajectory in the interior of it's past
lightcone. For a massless field, the radiation reaction force is local except
that the regulated Green's function $G_{\Lambda}^{R}$ smears out the radiation
reaction over the past trajectory expanded in a Taylor series around $z\left(
\tau\right)  $,as in Eq. (3.8).}%
\label{label1}%
\end{figure}

\begin{figure}[tbh]
\begin{center}
\epsfig{file=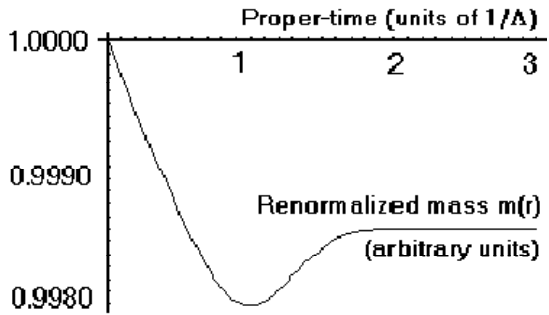, width=3 in}
\end{center}
\caption{The time-dependent renormalized mass $m\left(  r\right)  $ of the
particle plotted against the proper-time, $r=\tau-\tau_{i}$, ellapsed since
the intial time $t_{i}$ at which the factorized intial state of the particle
plus field is specified. The actual mass-shift depends on the cutoff
$\Lambda;$ the vertical mass units are arbitrary since they depend on the
unkown particle bare mass $m_{0}.$ The renormalization time-scale is
$\tau_{ren}\sim1/\Lambda.$}%
\label{label2}%
\end{figure}

\begin{figure}[tbh]
\begin{center}
\epsfig{file=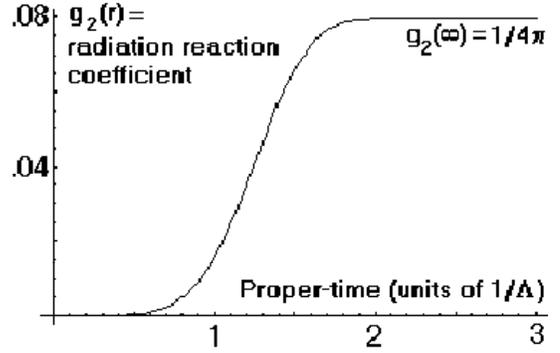, width=3 in}
\end{center}
\caption{The time-dependent coefficient $g_{2}\left(  r\right)  $ that
determines the radiation reaction (RR) plotted against the ellapsed
proper-time, $r=\tau-\tau_{i}$, since the intial time $t_{i} $ at which the
factorized intial state of the particle plus field is specified. The radiation
reaction vanishes at $r=0$, but quickly builds to the asymptotic value
familiar from the Abraham-Lorentz-Dirac equation on a timescale 1/$\Lambda.$}%
\label{label3}%
\end{figure}

\end{document}